\documentclass[10pt,journal,final]{IEEEtran}
\usepackage{mathrsfs}
\usepackage{graphicx}
\usepackage{float}
\usepackage{booktabs}
\usepackage{hyperref}
\usepackage{cite}
\usepackage{microtype}

\usepackage{enumitem}
\usepackage{color}
\usepackage{ulem}
\usepackage{psfrag}
\usepackage{subfigure}
\usepackage{amssymb}
\usepackage{amsthm}
\usepackage{setspace}
\usepackage{epsfig}
\usepackage{pifont}
\usepackage{amsmath}
\usepackage{array}
\newcommand{\PreserveBackslash}[1]{\let\temp=\\#1\let\\=\temp}
\newcolumntype{C}[1]{>{\PreserveBackslash\centering}p{#1}}
\newcolumntype{R}[1]{>{\PreserveBackslash\raggedleft}p{#1}}
\newcolumntype{L}[1]{>{\PreserveBackslash\raggedright}p{#1}}
\usepackage{multicol}
\usepackage{multirow}
\usepackage{diagbox}
\usepackage{pifont}
\usepackage{indentfirst}
\usepackage{amsfonts}
\usepackage{algorithm}
\floatname{algorithm}{Procedure}
\usepackage{algorithmicx}
\usepackage{algpseudocode}
\usepackage{fancyhdr}
\usepackage{amscd}
\usepackage[cmyk]{xcolor}
\usepackage{bm}
\usepackage{fancyhdr}
\setlist{itemsep=0pt,parsep=0pt}

\newtheorem{proposition}{Proposition}
\newtheorem{definition}{Definition}

\newtheorem{lemma}{Lemma}
\newtheorem{remark}{Remark}

\renewcommand{\arraystretch}{0.9} 
\allowdisplaybreaks[4]
\IEEEoverridecommandlockouts

\usepackage{caption}
\begin{document}
	\title{When Attention is Beneficial for Learning Wireless Resource Allocation Efficiently?}
	
	\author{
		\IEEEauthorblockN{Jia Guo and Chenyang Yang\vspace{-10mm}}
		%
	}
	\maketitle
	\setcounter{page}{1}
	\thispagestyle{headings}

	\begin{abstract}
       Owing to the use of attention mechanism to leverage the dependency across tokens, Transformers are efficient for natural language processing. By harnessing permutation properties broadly exist in resource allocation policies, each mapping measurable environmental parameters (e.g., channel matrix) to optimized variables (e.g., precoding matrix), graph neural networks (GNNs) are promising for learning these policies efficiently in terms of scalability and generalizability. To reap the benefits of both architectures, there is a recent trend of incorporating attention mechanism with GNNs for learning wireless policies. Nevertheless, \emph{is the attention mechanism really needed for resource allocation?}
       In this paper, we strive to answer this question by analyzing the structures of functions defined on sets and numerical algorithms, given that the permutation properties of wireless policies are induced by the involved sets (say user set). In particular, we prove that the permutation equivariant functions  on a single set can be recursively expressed by two types of functions: one involves attention, and the other does not. We proceed to re-express the numerical algorithms for optimizing several representative resource allocation problems in recursive forms. We find that \emph{when interference (say multi-user or inter-data stream interference) is not reflected in the measurable parameters of a policy, attention needs to be used to model the interference.} With the insight, we establish a framework of designing GNNs by aligning with the structures. By taking reconfigurable intelligent surface-aided hybrid precoding as an example, the learning efficiency of the proposed GNN is validated via simulations.
        	
	\end{abstract}
	
	\vspace{-4mm}
	\section{Introduction}\label{sec:intro}	\vspace{-1mm}
	Transformers have become a backbone architecture of large language models \cite{achiam2023gpt}, and have been applied to many fields such as computer vision. Encouraged by the success, recent works strive to introduce Transformers into wireless communications \cite{Transformer_MIMO_Semantic_WCM2023,LLM_Comm_WCM2024}. The core of Transformers is the attention mechanism, which can capture the dependence among tokens (say words for machine translation tasks) \cite{Transformer}. For channel acquisition tasks such as channel estimation \cite{Transformer_chl_est_TVT2024,trans_CE_payless_TWC} and channel prediction \cite{CP1,Transformer_chl_pre_TWC2024}, it is natural to adopt Transformers because they can leverage temporal and spatial correlations among~channels.
	
	Transformers have also been tailored for resource allocation tasks such as power allocation and precoding \cite{li2024hpe,Transformer_PA_WCL2025}. Yet it was found in  \cite{DYX} that the Transformer is inefficient for optimizing precoding, due to not exploiting the permutation equivariance (PE) properties of precoding policies, each being a mapping from channel matrices to precoding matrices.

More broadly, a resource allocation policy is the mapping from the \emph{measurable environmental parameters} (called \emph{environmental parameters} for short) to the optimized variables of a problem.
	The PE property of a policy come from the \emph{sets} involved in a resource allocation problem, say user set. When the ordering of users does not affect the objective function and constraints, the policy obtained from the problem is PE to users \cite{LSJ_MultiDim_GNN_2022,SCJ20ranking}. For instance, in the problem of optimizing precoding in multi-user multi-input-single-output
	(MU-MISO) systems, the antennas at the base station (BS) and the users constitute two sets. The resulting policy is the mapping from the feature of the sets (i.e., channel matrix) to another feature of these sets (i.e., precoding matrix) \cite{SCJ20ranking}, which is with two-dimensional (2D) input and output.
	Each set corresponds to a \emph{dimension}, hence the  policy on the antenna and user sets is with antenna- and user-dimensions, and exhibits the 2D-PE property \cite{ZBC_WCNC}.
	
	The elements in a set can be \emph{arbitrarily} or \emph{nestedly} permuted, and the elements in different sets can be independently or jointly permuted, depending on system settings \cite{LSJ_MultiDim_GNN_2022}.
	For example, the users in a single-cell system can be arbitrarily permuted, while the users in a multi-cell system can be nestedly permuted since the users in each cell should be permuted together with cells. The permutability of elements in sets of a problem affects the permutation property of the resulting policy. The permutation properties form a broader class of properties including PE.
	
	As an architecture that can harness PE properties, graph neural networks (GNNs) have been demonstrated capable of learning resource allocation policies more efficiently than fully-connected neural networks (FNNs) and convolutional neural networks (CNNs) in the following two aspects.
	\begin{enumerate}
		\item Better \emph{size generalizability}\cite{Eisen2020,SYF,GNN-PA-CellFree-TWC2024,Distributed_GNN_BF_RIS_MU_TWC2025}, indicating that GNNs can perform well under unseen problem scales (say the number of users) without re-training,
		\item Better \emph{scalability} \cite{lee2020wireless,GJ_TWC_GNN,GNN-PC-CellFree-TWC2024}, indicating that GNNs can be trained with low complexity for large-scale problems.
	\end{enumerate}
	
	It has been noticed that the advantages of GNNs stem from leveraging permutation properties widely exist in resource allocation policies \cite{Eisen2020,SYF, GJ_TWC_GNN,LSJ_MultiDim_GNN_2022}, especially when they are integrated with attention mechanism \cite{GJ-TMLCN,GNN_BF_MUMISO_SE_TWC2024}.
	
%
%

	\vspace{-3mm}
	\subsection{Related Works}\vspace{-1mm}

    \subsubsection{DNNs for resource allocation with permutation property}

GNNs can be designed to leverage the permutation properties. For learning a policy efficiently with a GNN, the input-output (I-O) relation of the GNN should exhibit the same permutation property as the policy. The permutation property satisfied by a GNN depends on the constructed graph (e.g., types of vertices and edges) and the GNN architecture. Most existing works first design the graphs and architectures of GNNs and then prove that the designed GNNs over the graphs are with desired permutation properties \cite{GJ_TWC_GNN,ZBC_WCNC,LY_GNN_RA_TWC2024,Distributed_GNN_BF_RIS_MU_TWC2025}. With this approach, the permutability induced by some sets of a problem is often overlooked, causing the mismatched property of a GNN with a policy, which leads to the non-generalizability and non-scalability to the corresponding dimensions \cite{GJ_TWC_GNN}. To exhibit matched permutation property to a policy, a framework of identifying sets from a problem, modeling graphs and designing parameter sharing of GNNs was proposed in \cite{LSJ_MultiDim_GNN_2022}.

Permutation equivariance neural networks (PENNs) can satisfy permutation properties by introducing parameter sharing into FNNs  \cite{PEandParamShare2017,zaheer2017deep}, which can be regarded as a special GNN that learns over a complete graph. In \cite{GJ2020ICC}, a PENN was designed for optimizing predictive resource allocation.

    It is widely believed that GNNs are naturally size generalizable and scalable, because the sizes of the weight matrices are independent of problem scales due to parameter sharing \cite{Eisen2020, SYF, LY_GNN_RA_TWC2024}. However, it has recently been found that both abilities of a GNN are affected by the processors, which are parameterized functions in GNN for extracting information from the representations of neighbouring vertices or edges \cite{GJ-TMLCN}. The commonly used processors are linear processors \cite{GJ_TWC_GNN,Eisen2020} and FNNs \cite{Distributed_GNN_BF_RIS_MU_TWC2025,GNN_BF_RIS_MU_TWC2025}.
    Empirical evaluations show that a GNN with linear or FNN processors is size generalizable  for power control/allocation or link scheduling \cite{GJ_TWC_GNN, Eisen2020, lee2020wireless}. For spectral efficiency (SE)-maximal MU-MISO precoding, by only learning a power allocation policy with a GNN and then recovering the precoding matrix with the optimal solution structure, the GNN with linear processor is also size generalizable \cite{Bipartie_GNN_BF_TWC2023}.
    For precoding problems without the structure, the precoding policy needs to be learned, and the GNNs with linear processors is neither size generalizable nor scalable \cite{ZBC_WCNC}. As analyzed in \cite{GJ-TMLCN,GJ-MLSP}, to enable the generalizability of GNN for learning MU-MISO and multi-user multi-input-multi-output (MU-MIMO) precoding, the processor should model the correlation among channels of users.

    \subsubsection{DNNs for resource allocation with attention mechanism}
    The Transformer proposed for neural language processing is with an encoder-decoder architecture, and the order of tokens is distinguished with a positional encoding. For the tasks of resource allocation, encoder-only Transformers without positional encoding have been adopted. For example, a multi-group multicast precoding was optimized with such a Transformer in \cite{li2024hpe}. The Transformer was actually used to learn power allocation, and the precoding matrix was recovered from the allocated power with an optimal solution structure.  A power allocation policy in a cell-free system was learned with the Transformer in \cite{Transformer_PA_WCL2025}. The bandwidth of subbands, precoding, and reflection coefficients in a reconfigurable intelligent surface (RIS)-aided multi-antenna system were jointly optimized by the Transformer in \cite{mehrabian2024joint}.

    Another DNN with attention mechanism is the graph attention network (GAT) \cite{GAT_2017}, which has been adopted for learning wireless policies. For example, a precoding policy in a MU-MISO system was learned by a GAT in \cite{GNN_BF_MUMISO_SE_TWC2024} over a complete graph with user vertices, where the processor is with attention to model the correlation of channel vectors of users. To optimize precoding in multi-user RIS systems, a GAT-based DNN was designed in \cite{GAT_RIS_MUMISO_SE_TVT2024}.

    While both the Transformer and GAT are generalizable to the number of users \cite{li2024hpe,GNN_BF_MUMISO_SE_TWC2024}, they are still not efficient since the equivariance to the permutations of antennas is neglected.

\vspace{-3.5mm}\subsection{Motivations and Contributions}\vspace{-0.5mm}
Resource allocation problems (e.g., multi-cell multi-user power allocation \cite{GJ_TWC_GNN} and RIS-aided resource allocation \cite{mehrabian2024joint,RIS_BCD}) uauslly consist of multiple sets. For learning the resulting policies with size generalizability and scalability, high-dimensional properties regarding the permutability of elements in all the sets should be exploited, and processors with attention may need to be used. However, computing attention scores incurs quadratic complexity, which is higher than the linear or FNN processors with linear complexity.

When learning wireless policies, there are various ways of computing attention scores and not all of them perform well. Again taking learning MU-MISO precoding as an example, the attention between two channel coefficients or channel vectors of users can be computed, while only the second way helps improve learning efficiency \cite{GNN_BF_MUMISO_SE_TWC2024,GJ-TMLCN}. Existing GNNs or Transformers either consider only one set (say user set) while neglecting other sets (say antenna set) \cite{li2024hpe, GNN_BF_MUMISO_SE_TWC2024}, or incorporate attention mechanism in all dimensions \cite{GJ-TMLCN}.
So far, the DNNs with both permutation properties and attention mechanism are still designed with trial-and-error. This is because attention mechanism seems irrelevant to the permutation properties.
To guide the design of a DNN with low inference complexity and high learning efficiency, it is important to answer the following two questions.
%
({\bf Q1}) When attention mechanism is beneficial?
({\bf Q2}) How to satisfy high-dimensional permutation properties meanwhile only introducing the attention mechanism into necessary dimensions?

Since attention is capable of capturing  correlations (i.e., dependence),
it should also be able to reflect interference such as multi-user interference (MUI) or inter-stream interference (IDI) caused by the correlation of channels of different entities.
To analyze the role of attention in resource allocation, we consider two categories of problems, which are divided according to the scenarios with and without interference. In the two aspects of learning efficiency, we focus on size generalizability, because a size-generalizable DNN does not need to be re-trained for large-scale problems.

    The major contributions are summarized as follows.
    \begin{itemize}
        \item We reveal the structures of functions on a single set, which are with or without attention. We show that the function on multiple sets can be constructed by one or two of these functions in a recursive manner, where the permutability induced by a set is satisfied in each recursion. Hence, these functions behave like  ``basis functions''. 
        \item We find that attention exists for modeling interference when interference is not reflected in the environmental parameters of a resource allocation policy, by expressing the iterative equations of numerical algorithms for solving several problems in recursive forms.
        \item We establish a framework for systematically incorporating attention with GNNs, which satisfy high-dimensional permutation properties recursively. The main design aspects, including the number of recursions and the  ``basis functions'' in each recursion, can be determined by analyzing a resource allocation problem. By taking learning a multi-cell RIS-aided precoding policy as an example, the size generalizability of the GNN designed by the proposed framework is evaluated via simulations.
    \end{itemize}

	The rest of the paper is organized as follows.
	In section \ref{sec:pe-funcs}, we introduce functions satisfying PE properties and reveal their structures. In section \ref{sec:design-gnns}, we re-express iterative equations of algorithms for different problems in recursive form, and summarize when attention is beneficial for resource allocation. In section \ref{sec:gnn-design-framework}, we propose the framework for designing GNNs with attention. Simulation results and concluding remarks are respectively provided in section \ref{sec:simulation-results} and \ref{sec:conclusions}.
	
	\emph{Notations}: Lower and uppercase letters respectively denote vectors and matrices.
$\mathbf{A}=[a_{ij}]^{M\times N}$ denotes a $M\times N$ matrix whose element on the $i$-th row and the $j$-th column is $a_{ij}$. $\vec{\mathbf{A}}$ denotes a tensor. ${\mathbb R}$ and $\mathbb C$  respectively represent the sets of real and complex numbers. $\odot$ denotes Hadmard product and $\otimes$ denotes Kronecker product. $\mathbf{0}_{M\times N}$ and $\mathbf{1}_{M\times N}$ are respectively $M\times N$ matrices with all the elements being $0$ and $1$. $\mathbf{I}_K$ stands for a $K\times K$ unit matrix. $\|\cdot\|$ and $\|\cdot\|_F$ denote two-norm and Frobenius norm, respectively. $f(\cdot):{\mathbb R}^{M}\mapsto {\mathbb R}^N$ represents a mapping from a vector of size $M$ to a vector of size $N$. $\pi(\cdot)$ denotes the permutation operation of a set.

    \vspace{-2mm}
\section{Structures of the Functions on Sets}\label{sec:pe-funcs}
	\vspace{-0.5mm}
    In this section, we define two types of sets and the resulting PE properties. Then, we reveal the structures of four functions with the PE properties (called PE functions) induced by a single set. Finally, we demonstrate that the PE functions defined on more than two sets can be constructed by these functions in a recursive manner.
    \vspace{-2mm}
\subsection{Preliminary}	\vspace{-0.5mm}
    We consider two types of sets, normal sets and nested sets. A normal set is a set where the order of elements in the set does not matter. A nested set is a set consisting of multiple subsets. The order of subsets and the order of elements in each subset do not matter. The PE properties of functions on sets depend on the number, types, and relation of the sets.

\subsubsection{PE property induced by a normal set} Denote a function on a normal set \{1,$\cdots$, $K$\} as $\mathbf{y}=\phi(\mathbf{x})$, where $\mathbf{x}=[x_1,\cdots,x_K]^T$ and $\mathbf{y}=[y_1,\cdots,y_K]^T$ are respectively the input and output feature vectors of the set.\footnote{$x_k$ and $y_k$ are scalars, which can be replaced by vectors, matrices and tensors in the following PE functions.}

Denote $\mathbf{\Pi}$ as an arbitrary permutation matrix consisting of $0$ and $1$, which can change the order of $K$ elements of the set in $K!$ ways.
The function on the normal set satisfies an arbitrary PE (APE) property defined as follows \cite{zaheer2017deep}.
\begin{definition}
	The function $\mathbf{y}=\phi(\mathbf{x})$ satisfies APE property iff $\bm\Pi^T\mathbf{y}=\phi(\bm\Pi^T\mathbf{x})$, which is referred to as the APE function.
\end{definition}

\subsubsection{PE property induced by a nested set}
Denote a function on a nested set $\{\{{1_1},\cdots,{K_1}\},$ $\cdots,\{{1_M},\cdots,{K_M}\}\}$ as  $\mathbf{y}=\phi(\mathbf{x})$, where $\mathbf{x}=[\mathbf{x}_1^T,\cdots,\mathbf{x}_M^T]^T$ and $\mathbf{y}=[\mathbf{y}_1^T,\cdots,\mathbf{y}_M^T]^T$, $\mathbf{x}_m=[x_{1_m},\cdots,x_{K_m}]^T$ and $\mathbf{y}_m=[y_{1_m},\cdots,y_{K_m}]^T$.

Denote $\mathbf{\Omega}$ as a nested permutation matrix.
In particular,
\begin{equation}\label{eq:nest-perm-mat}
	\bm\Omega=(\mathbf{\Pi}_0\otimes\mathbf{I}_K)\mathsf{diag}(\mathbf{\Pi}_1,\cdots,\mathbf{\Pi}_M),
\end{equation}
where $\mathbf{\Pi}_0$ is an arbitrary permutation matrix that can re-order the sub-vectors of $\mathbf{x}$ and $\mathbf{y}$ in $M!$ ways, and $\mathbf{\Pi}_m$ is an arbitrary permutation matrix enabling to re-order the elements in  $\mathbf{x}_m$ and $\mathbf{y}_m, m=1,\cdots,M$  in $K!$ ways.

\begin{figure}[!htb]
	\centering
	\vspace{-1mm}
	\includegraphics[width=0.5\linewidth]{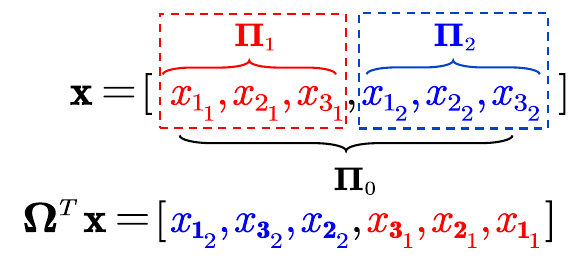}
	\vspace{-2mm}
	\caption{An example of permuted vector, $M=2,K=3$, where both the sub-vectors and the elements of each sub-vector are re-ordered.}
	\label{fig:fig-perm-mat}
	\vspace{-0.2mm}
\end{figure}

In Fig. \ref{fig:fig-perm-mat}, we illustrate how the nested permutation matrix changes the order of elements in vector $\mathbf{x}$.

The function on the nested set satisfies a nested PE (NPE) property as follows  \cite{PEandParamShare2017}.
\vspace{-1mm}
\begin{definition} \label{def:npe}
	The function $\mathbf{y}=\phi(\mathbf{x})$ satisfies NPE property iff
	$\bm\Omega^T\mathbf{y}=\phi(\bm\Omega^T\mathbf{x})$, which is referred to as the NPE function.
\end{definition}
\vspace{-1mm}
 \subsubsection{PE property induced by two sets} While the two sets can either be normal sets, nested sets or their combination, we only define the functions on two normal sets for easy exposition.
    A function on two normal sets satisfies 2D-PE property if the elements in the two sets can be permuted independently, or joint-PE property if the elements in the two sets can only be permuted jointly (i.e., dependently).
    \vspace{-1mm}
    \begin{definition}\label{def:2d-pe}
        A function $\mathbf{Y}=\phi(\mathbf{X})$ mapping $\mathbf{X}=[\mathbf{x}_1,\cdots,\mathbf{x}_K]$ to $\mathbf{Y}=[\mathbf{y}_1,\cdots,\mathbf{y}_K]$ satisfies 2D-PE property iff $\bm\Pi_1^T\mathbf{Y}\bm\Pi_2=\phi(\bm\Pi_1^T\mathbf{X}\bm\Pi_2)$, which is called 2D-PE function.
    \end{definition}
    \vspace{-2mm}
	\begin{definition}\label{def:joint-pe}
		A function $\mathbf{Y}=\phi(\mathbf{X})$ or $\mathbf{y}=\phi(\mathbf{X})$ satisfies joint-PE property iff $\bm\Pi^T\mathbf{Y}\bm\Pi=\phi(\bm\Pi^T\mathbf{X}\bm\Pi)$ or $\bm\Pi^T\mathbf{y}=\phi(\bm\Pi^T\mathbf{X}\bm\Pi)$, which is called joint-PE function.
	\end{definition}

\vspace{-4mm}
\subsection{Structures of the Functions on a Single Set}
\subsubsection{Two APE functions on a normal set}
As proved in \cite{PE_decomposition}, every (say the $k$-th) output variable of any function $\phi(\mathbf{x})$ satisfying the APE property
 can be expressed as,
\begin{equation}
	y_k = f_{\rm I}(x_k, \textstyle\sum_{j=1,j\neq k}^K q_{\rm I}(x_j)), \label{eq:1d-pe-I}
\end{equation}
where $q_{\rm I}(\cdot):{\mathbb R}\mapsto{\mathbb R}^{K}$, $f_{\rm I}(\cdot):{\mathbb R}^{K+1}\mapsto {\mathbb R}$, which are respectively called processor and combiner in the sequel.

It is not hard to prove that a function whose output variable can be expressed as follows also satisfies the APE property,
\begin{equation}
	y_k = f_{\rm II}(x_k, \textstyle\sum_{j=1,j\neq k}^K q_{\rm II}(x_k,x_j)), \label{eq:1d-pe-II}
\end{equation}
where $q_{\rm II}(\cdot):{\mathbb R}\mapsto{\mathbb R}$, $f_{\rm II}(\cdot):{\mathbb R}\mapsto {\mathbb R}$, i.e., the input and output sizes of the processor and combiner are independent of $K$, which are different from $f_{\rm I}(\cdot)$ and $q_{\rm I}(\cdot)$.

This does not contradict the statement that every output variable of any function with APE property can be expressed in \eqref{eq:1d-pe-I}, because of a fact proved in the following proposition.

\begin{proposition}\label{prop:APE}
	\eqref{eq:1d-pe-II} can be re-expressed in the form in \eqref{eq:1d-pe-I}.
	
	\begin{IEEEproof}
		See Appendix \ref{Appendix:re-express}.
	\end{IEEEproof}
\end{proposition}

The two APE functions with the structures in \eqref{eq:1d-pe-I} and \eqref{eq:1d-pe-II} are respectively referred to as \emph{APE function-I} and \emph{APE function-II}, which differ in their processors. Specifically, $q_{\rm I}(\cdot)$ is a function of only $x_j$, while $q_{\rm II}(\cdot)$ is a function of both $x_k$ and $x_j$ and hence can model an ``attention'' between the two variables. In the following,
$q_{\rm I}(\cdot)$ and $q_{\rm II}(\cdot)$ are respectively referred to as \emph{ordinary processor} and \emph{attention processor}.


\subsubsection{Two NPE functions on a nested set} We next show the structures of the NPE functions.

\begin{proposition}\label{prop:npe}
Every (say the $k_m$-th) output variable of any function $\phi(\mathbf{x})$ satisfying the NPE property can be expressed as follows,
	\begin{align}\label{eq:nest-pe-I}
		&y_{k_m} = f_{\rm I}\Big(x_{k_m}, \textstyle\sum_{j=1,j\neq k}^K q_{\rm I,1}(x_{j_m}), \notag\\
		&\hspace{15mm}\textstyle\sum_{i=1,i\neq m}^M q_{\rm I, 2}\big(\sum_{j=1}^K q_{\rm I,3}(x_{j_i})\big)\Big),
	\end{align}
	where $f_{\rm I}(\cdot): {\mathbb R}^{1+K+MK}\mapsto{\mathbb R}$ is a combiner, and
	$q_{\rm I,1}(\cdot):{\mathbb R}\mapsto{\mathbb R}^K$,
	$q_{\rm I, 2}(\cdot): {\mathbb R}^{K+1}\mapsto {\mathbb R}^{MK}$, and
	$q_{\rm I, 3}(\cdot):{\mathbb R}\mapsto {\mathbb R}^{K+1}$ are ordinary processors.
	\begin{IEEEproof}
		See Appendix \ref{appendix:proof-npe}.
	\end{IEEEproof}
\end{proposition}
Again, it is not hard to prove that a function whose output variable
can be expressed as follows  satisfies the NPE property,
\begin{align}\label{eq:nest-pe-II}
	&y_{k_m} = f_{\rm II}\Big(x_{k_m}, \textstyle\sum_{j=1,j\neq k}^K q_{\rm II,1}(x_{k_m},x_{j_m}), \notag\\
	&\hspace{15mm}\textstyle\sum_{i=1,i\neq m}^M q_{\rm II, 2}\big(\sum_{j=1}^K q_{\rm II,3}(x_{k_m},x_{j_i})\big)\Big),
\end{align}
where the input and output sizes of the combiner $f_{\rm II}(\cdot)$, the ordinary processor  $q_{\rm II, 2}(\cdot)$, as well as the attention processors $q_{\rm II,1}(\cdot)$ and $q_{\rm II, 3}(\cdot)$ are independent of $K$.

With a similar proof to Appendix \ref{Appendix:re-express}, we can show that \eqref{eq:nest-pe-II} can be expressed in the form of \eqref{eq:nest-pe-I}.

The functions  with the structures in \eqref{eq:nest-pe-I} and \eqref{eq:nest-pe-II} are respectively referred to as \emph{NPE function-I} and \emph{NPE function-II}.



\vspace{-1mm}\begin{remark}\emph{
	Proposition \ref{prop:npe} can be extended to functions defined on a more general nested set with $N_{\sf E}$ tiers. A general nested set consists of multiple subsets in the first tier, and each of them also consists of multiple subsets in the second tier. This nesting continues in the same manner for $N_{\sf E}$ times. In the last tier of nesting, each subset consists of multiple elements. Any function on the nest set with $N_{\sf E}>2$ tiers can be expressed in a form similar to \eqref{eq:nest-pe-I}, but with $1+\cdots+N_{\sf E}$ summations and processors. For notational simplicity, the ``nested set'' and ``NPE function'' respectively refer to the sets and functions with $N_{\sf E}=2$ in the rest of this paper.}
\end{remark}\vspace{-1mm}

We refer to the functions  with the structures in \eqref{eq:1d-pe-I} $\sim$ \eqref{eq:nest-pe-II} as \emph{one-set PE functions}.
They can be divided into two types based on the satisfied PE properties. Each type contains two functions, depending on whether attention processor is used.

The one-set PE functions behave like basis functions, because the functions defined on multiple sets can be constructed by these functions as to be shown in what follows.
	
	\vspace{-2mm}
	\subsection{Structures of the Functions on Multiple Sets}\label{sec:pe-func-multiset}	\vspace{-0.1mm}
    The functions defined on multiple sets are called multi-set PE functions.
We first take functions defined on two sets as examples to demonstrated that multi-set PE function can be constructed by the one-set PE functions in a recursive manner.

    \subsubsection{Functions on two sets}
    The permutation properties of the functions on two sets depend on the type of each set and the relation between the sets. Again, we take the functions defined on normal sets as examples, then a two-set PE function either satisfies the 2D-PE or satisfies the joint-PE property.

    The following proposition shows that a function expressed by the APE functions recursively has the 2D-PE property.
    \begin{proposition}\label{prop:2d-pe}\emph{
        If every (say the $k$-th) output variable of the function $\mathbf{Y}=\phi(\mathbf{X})$ can be expressed in the following form, then it satisfies the 2D-PE property. \\
        \uline{\emph{First recursion}}: The mapping from $\mathbf{x}_k, k=1,\cdots, K$ to $\mathbf{y}_k$ is an APE function-I as,
        \begin{align}
            \mathbf{y}_k=f\big(\mathbf{x}_k, \textstyle\sum_{j=1,j\neq k}^K q(\mathbf{x}_j)\big), \label{eq:2d-pe-r1}
        \end{align}
        where $\mathbf{x}_k=[x_{1k},\cdots,x_{Nk}]^T, \mathbf{y}_k=[y_{1k},\cdots,y_{Nk}]^T$.
        \uline{\emph{Second recursion}}: $f(\cdot)$ \!and \!$q(\cdot)$ \!in \!\eqref{eq:2d-pe-r1} \!are \!APE functions-I \!as,
        \begin{align}
            f(\cdot):~ & y_{nk} = f_1\big(\mathbf{x}_{nk}', \textstyle\sum_{i=1,i\neq n}^N q_1(\mathbf{x}_{ik}')\big), \label{eq:2d-pe-r2-1}\\
            q(\cdot):~ & z_{nk} = f_2\big(x_{nk}, \textstyle\sum_{i=1,i\neq n}^N q_2(x_{ik})\big), \label{eq:2d-pe-r2}
        \end{align}
        where $z_{nk}$ is the $n$-th element of a vector $\mathbf{z}_k \triangleq q(\mathbf{x}_k)$, and $\mathbf{x}_{nk}' \triangleq [x_{nk},\sum_{j=1,j\neq k}^K z_{nj}]^T$. }
        \begin{IEEEproof}
        	See Appendix \ref{proof:2d-pe}.
        \end{IEEEproof}
    \end{proposition}

    With similar proof in Appendix \ref{proof:2d-pe}, we can prove that $\mathbf{Y}=\phi(\mathbf{X})$ still satisfies  the 2D-PE property when the APE functions-I in \eqref{eq:2d-pe-r1}, \eqref{eq:2d-pe-r2-1} and/or \eqref{eq:2d-pe-r2} are replaced by APE-functions-II, and when the APE functions in the two recursions are different.

    As shown in Proposition \ref{prop:2d-pe}, the input of $q(\cdot)$ in the first recursion is a vector, while the input of the processor $q_1(\cdot)$ in the second recursion is a scalar. This means that the input dimensions of the processors decrease with recursions. Hence, if the APE function-I in the first or second recursion becomes an APE function-II, then the attention processor reflects the correlation between two vectors or two scalars.

        The following proposition indicates that a joint-PE function can be constructed by a composite function of a 2D-PE function and an output function $f_{\rm out}(\cdot)$.

    \begin{proposition}\label{prop:joint-pe}\emph{
        The mapping from $\mathbf{X}\in{\mathbb R}^{K\times K}$ to $\mathbf{Y}\in{\mathbb R}^{K\times K}$ satisfies the joint-PE property if it is in the form of $\mathbf{Y}=\phi(\mathbf{X})\odot\mathbf{I}_K + \mathbf{X}\odot (1-\mathbf{I}_K) \triangleq f_{\rm out}(\phi(\mathbf{X}),\mathbf{X})$, where $\phi(\mathbf{X})$ is a 2D-PE function.}
        \begin{IEEEproof}
        	See Appendix \ref{proof:joint-pe}.
        \end{IEEEproof}
    \end{proposition}

    By cascading with $f_{\rm out}(\cdot)$, the diagonal elements of $\mathbf{Y}$ are the same as the diagonal elements in the output of $\phi(\mathbf{X})$, while the non-diagonal elements of $\mathbf{Y}$ are the same as those in $\mathbf{X}$.

   \begin{remark} \emph{
   	We can also construct two-set PE functions defined on other types of sets recursively. Specifically, if one of the two sets is a nested set, then the one-set PE function(s) in a recursion are NPE functions-I/II.}
   \end{remark}

    \subsubsection{Functions on more than two sets}
    For the PE functions defined on $S>2$ sets, the input $\vec{\mathbf{X}} \in {\mathbb R}^{N_1\times\cdots\times N_S}$ and the output $\vec{\mathbf{Y}} \in {\mathbb R}^{N_1\times\cdots\times N_S}$ of a $S$-set function are tensors with $S$ dimensions. The function can expressed in the form of $S$ recursions. There is only a single one-set PE function in the first recursion, and more one-set PE functions in the subsequent recursions.
    The one-set PE function(s) in the $s$-th recursion is equivariant to the permutations in the $s$-th dimension of $\vec{\mathbf{X}}$,
    whose structure depends on the type of the $s$-th set. When the set is  a normal or nested set, the one-set PE function(s) are APE or NPE function(s).
If the elements in a set can only be jointly permuted, then $f_{\rm out}(\cdot)$ is cascaded after $S$ recursions.

	\section{Structures of Numerical Algorithms for Several Wireless Problems}\label{sec:design-gnns}
	In this section, we uncover the structures of the numerical algorithms for several problems involving sets. We show that the algorithms can be re-expressed in the recursive forms with one-set PE functions. From the structure of the re-expressed iterative equation (RIE) of each problem, we can observe when attention appears and how attention is computed.

    We consider problems in the scenarios with and without interference. We acknowledge that the considered problem in the interference-free scenario is too simple, which makes using learning-based methods unnecessary. Nonetheless, it facilitates the analysis of the role of attention in resource allocation.
	
	\vspace{-2mm}
	\subsection{Interference-Free Scenario}\label{subsec:interf-free}
	%
	Consider an orthogonal frequency division multiple access system, where a BS serves $K$ user equipments (UEs). We take the following bandwidth and power allocation problem as an example, which minimizes the total bandwidth required to satisfy the minimal data rate $s_0$,
	\begin{subequations}
		\begin{align}
			\textsf{P-B}:\min_{\substack{p_1,\cdots,p_K\\B_1,\cdots,B_K}}~& \textstyle\sum_{k=1}^K B_k,\\
			{\rm s.t.} ~& s_k\!\triangleq\! B_k\log_2(1+p_kg_k/(N_0B_k))\geq s_0, \label{eq:pb-1}\\
			& \textstyle\sum_{k=1}^K p_k\leq P_{\max}, p_k\geq 0, B_k\geq 0 \label{eq:pb-2},
		\end{align}
	\end{subequations}
	where $B_k, p_k$ and $g_k$ are respectively the allocated bandwidth, power and channel gain of the $k$-th UE (denoted as UE$_k$), and $N_0$ is the noise spectral density.

	Problem  \textsf{P-B} consists of a normal set, i.e., UE set.
	
	Denote the joint power and bandwidth allocation policy as  $(\mathbf{p}^\star,\mathbf{b}^\star)=\phi_\mathsf{B}(\mathbf{g})$, where $\mathbf{b}^\star=[B_1^\star,\cdots,B_K^\star]^T$, $\mathbf{p}^\star=[p_1^\star,\cdots,p_K^\star]$,  and $\mathbf{g}=[g_1,\cdots,g_K]$ are respectively the optimized bandwidth and power allocation vectors, and the channel vector.
	It is not hard to prove that the policy satisfies the following permutation property, $(\bm\Pi_{\sf UE}^T\mathbf{p}^*,\bm\Pi_{\sf UE}^T\mathbf{b}^*)=\phi_\mathsf{B}(\bm\Pi_{\sf UE}^T\mathbf{g})$.
	
	The global optimal solution of the problem can be found via a gradient descent (GD) algorithm because the problem is convex. The constraints of the problem can be satisfied by the Lagrange multiplier method \cite{convex-opt}. Denote the Lagrange function as $\mathcal{L}\triangleq\sum_{k=1}^K  B_k + \sum_{k=1}^K  \mu_k(s_0-s_k) + \lambda\big(\sum_{k=1}^K p_k-P_{\max}\big)$, where $\mu_k$ and $\lambda$ are Lagrange multipliers for satisfying \eqref{eq:pb-1} and \eqref{eq:pb-2}, respectively. After deriving its gradient with respect to the optimization variables and Lagrange multipliers, the iterative equation in the $(\ell+1)$-th iteration of the GD algorithm can be expressed as,
	%
	%
	\begin{subequations}
		\begin{align}
			p_k^{(\ell+1)} 
			&=\textstyle\Big[p_k^{(\ell)}-\frac{\mu_k^{(\ell)}g_k}{\ln 2}\frac{1}{N_0B_k+p_k^{(\ell)}g_k}+\lambda^{(\ell)}\Big]^+,\label{eq:iter-bp-1}\\
			B_k^{(\ell+1)}
			&=\textstyle\Big[B_k^{(\ell)}-1-\mu_k^{(\ell)}\log_2\Big(1+\frac{p_k^{(\ell)}g_k}{N_0B_k^{(\ell)}}\Big)\notag\\
			& \hspace{10mm} \textstyle+\mu_k^{(\ell)}\frac{p_k^{(\ell)}g_k}{B_k^{(\ell)}N_0+p_k^{(\ell)}g_k}\Big]^+,\label{eq:iter-bp-2}\\
			\mu_k^{(\ell+1)}
			&=\textstyle \Big[\mu_k^{(\ell)} + B_k^{(\ell)}\log_2\Big(1+\frac{p_k^{(\ell)}g_k}{N_0B_k^{(\ell)}}\Big)-s_0\Big]^+,\label{eq:iter-bp-3}\\
			\lambda^{(\ell+1)} 
			&=\textstyle \Big[\lambda^{(\ell)} + \sum_{j=1}^K p_j^{(\ell)}-P_{\max}\Big]^+,\label{eq:iter-bp-4}
		\end{align}
	\end{subequations}
	where $[x]^+\triangleq\max(x,0)$.

	
	\begin{proposition}\label{prop:reexp-power-bandw-allo}\emph{	
			The I-O relation of the equation in the $(\ell+1)$-th iteration of the GD algorithm can be re-expressed as an RIE with an APE function-I as follows,
			\begin{equation}\label{eq:pb}
				\mathbf{d}_k^{(\ell+1)} = f_{\mathsf{B}}\Big(\mathbf{d}_k^{(\ell)}, \textstyle\sum_{j=1,j\neq k}^K q_{\mathsf{B}}(\mathbf{d}_j^{(\ell)})\Big),
			\end{equation}
			where $\mathbf{d}_k^{(\ell)}=[p_k^{(\ell)},B_k^{(\ell)},\mu_k^{(\ell)},\lambda^{(\ell)},g_k]^T$ is the ``\emph{representation vector}'' of the $k$-th user with five elements,
			$f_{\mathsf{B}}(\cdot)$ is provided in \eqref{eq:fb-expression}, and $q_{\mathsf{B}}(\mathbf{d}_j^{(\ell)})$ is a linear function provided in \eqref{eq:qb-expression}.
			\emph}
		\begin{IEEEproof}
			See Appendix \ref{proof:prop:reexp-power-bandw-allo}.
		\end{IEEEproof}
	\end{proposition}
	
	
As shown in the appendix, both the input and output sizes of $f_{\sf B}(\cdot)$ and $q_{\sf B}(\cdot)$ are independent from $K$. Hence, they are \emph{UE-size invariant functions}. 

The updated variables of all the users can be written as a matrix $\mathbf{D}^{(\ell)}=[\mathbf{d}_1^{(\ell)},\cdots,\mathbf{d}_K^{(\ell)}]\in {\mathbb R}^{5\times K}$, where the first dimension with five elements is the ``\emph{representation dimension}'', and the second dimension is the UE-dimension corresponding to the user set. Since the representation dimension is not relevant to the set in the problem (and hence the permutation property of the policy), we omit this dimension when expressing the updated variables in the sequel.

\vspace{1mm}\textbf{Observation 1}: As shown from the RIE in \eqref{eq:pb}, $q_{\mathsf{B}}(\cdot)$ is an ordinary linear processor instead of an attention processor.

\vspace{-3mm}
\subsection{Interference Scenarios} \label{subsec:interf}
\vspace{-0.5mm}
We consider two kinds of problems, relying on whether interference information is reflected in environmental parameters.

\vspace{0.1mm}\subsubsection{{Interference is not reflected in environmental parameters}}\label{sec:mu-miso-precoding}
To analyze the dependence of the structure of algorithms on the sets and interference in a problem,
we start with an example of optimizing precoding in MU-MISO systems with MUI and then extend to MU-MIMO systems with both MUI and IDI.

\textbf{Example 1 (Precoding for MU-MISO system)}: Consider a system where a BS with $N_{\sf B}$ antennas serves $K$ UEs each with a single antenna (AN). The precoding matrix can be optimized, say by maximizing the SE from the following problem,
\begin{subequations}
	\begin{align}
		\hspace{-2mm}\textsf{P-S}:
		\max_{\mathbf{W}}~~ & \sum_{k=1}^K \log_2 \left(1+\frac{|{\mathbf{h}}^{H}_k {\mathbf{w}}_{k}|^2}{\sum_{i=1,i\neq k}^K  |{\mathbf{h}}^{H}_k {\mathbf{w}}_{i}|^2+\sigma_0^2} \right) \label{eq:bb-object} \\
		{\mathrm{s.t.}} ~~
		&  \|{\mathbf{W}} \|^2_F \leq P_{\max},\label{eq:bb-constraint}
	\end{align}
\end{subequations}
where $\mathbf{w}_{k}=[w_{1k},\cdots,w_{N_{\sf B}k}]^T$ and $\mathbf{h}_k=[h_{1k},\cdots,h_{N_{\sf B}k}]^T$ are respectively the precoding and channel vectors of UE$_k$, $\mathbf{W}=[\mathbf{w}_{1},\cdots,\mathbf{w}_{K}]\in{\mathbb C}^{N_{\sf B}\times K}$, and $\sigma_0^2$ is the noise power.

The \!sets \!in \!the \!problem \!and \!their \!relation \!are \!as \!follows \!\cite{LSJ_MultiDim_GNN_2022},
\begin{itemize}
	\item \emph{Type of sets}: The UE set and AN set are normal sets.
	\item \emph{Relation among sets}: The elements in the two sets can be permuted independently.
\end{itemize}

Denote the precoding policy as ${\mathbf{W}}^\star=\phi_{\mathsf{S}}(\mathbf{H})$, where $\mathbf{W}^\star\in{\mathbb C}^{N_{\sf B}\times K}$ is the optimized precoding matrix, and $\mathbf{H}=[\mathbf{h}_1,\cdots,\mathbf{h}_K] \in{\mathbb C}^{N_{\sf B}\times K}$, both matrices are with AN-dimension and UE-dimension. In this policy, MUI (i.e., $|\mathbf{h}_k^{H}\mathbf{w}_{i}|^2$ in \eqref{eq:bb-object}) is not reflected in the environmental parameters (i.e., $\mathbf{H}$).

The policy satisfies the 2D-PE property \cite{ZBC_WCNC}.



Problem \textsf{P-S} is non-convex and a suboptimal solution can be found by the WMMSE algorithm \cite{WMMSE2011Shi}. After omitting the representation dimension for notational simplicity, the updated variables in the $\ell$-th iteration and the channel vector can be expressed as a matrix $\mathbf{D}^{(\ell+1)}=[\mathbf{d}_1^{(\ell+1)},\cdots,\mathbf{d}_{K}^{(\ell+1)}]\in{\mathbb C}^{N_{\sf B}\times K}$ with AN- and UE-dimensions, where representation
vector
\begin{align}\label{eq:upd-vars}
\mathbf{d}_k^{(\ell)}=\mathsf{concat}(\mathbf{w}_{k}^{(\ell)},u_k^{(\ell)},z_k^{(\ell)},\mathbf{h}_k),
\end{align}
$\mathbf{w}_{k}^{(\ell)}$ and $u_k^{(\ell)},z_k^{(\ell)}$ are respectively the updated precoding vector and intermediate variables for UE$_k$ in the $\ell$-th iteration, and $\mathsf{concat}(\cdot)$ is used to concatenate variables with different dimensions
(detailed definition was provided in \cite{GJ-TMLCN}).
\begin{proposition}\label{prop:reexp-miso-prec}
	\emph{The iterative equation of the WMMSE algorithm can be re-expressed as an RIE with two recursions. \\
		\noindent \underline{\emph{First recursion}}: The I-O relation of the RIE is an APE function-II along the UE-dimension, i.e.,
		\begin{equation}\label{eq:precoding}
			\mathbf{d}_k^{(\ell+1)} = f_{\mathsf{S}}\Big(\mathbf{d}_k^{(\ell)},\textstyle\sum_{j=1,j\neq k}^K q_{\mathsf{S}}(\mathbf{d}_k^{(\ell)},\mathbf{d}_j^{(\ell)})\Big),
		\end{equation}
        where $f_{\sf S}(\cdot)$ and $q_{\sf S}(\cdot)$ are UE-size invariant.\\
		\noindent \underline{\emph{Second recursion}}:
		$\mathbf{y}=f_{\mathsf{S}}(\mathbf{x})$ and $\mathbf{y}=q_{\mathsf{S}}(\mathbf{x})$ are APE functions-I along the AN-dimension, where the combiners and
processors are independent of $N_{\sf B}$ and $K$, hence are UE- and
AN-size invariant functions.
	}
	\begin{IEEEproof}
		The proof is similar to that in Appendix \ref{proof:prop:reexp-power-bandw-allo}, hence is omitted for conciseness.
	\end{IEEEproof}
\end{proposition}

\begin{figure}[!htb]
	\centering
        \vspace{-1mm}
	\includegraphics[width=.65\linewidth]{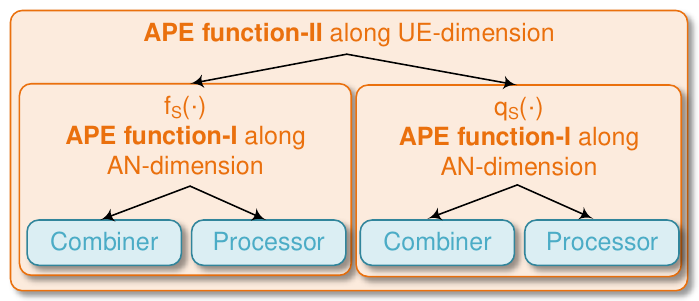}
	\vspace{-1mm}
	\caption{Structure of the RIE of the WMMSE algorithm for problem \textsf{P-S}, where different colors indicate different recursions.}
	\vspace{-1mm}
	\label{fig:rie-ps}
\end{figure}

The structure of the RIE is illustrated in Fig. \ref{fig:rie-ps}.

\textbf{Observation 2}: As shown in Fig. \ref{fig:rie-ps}, attention processor only appears in the one-set PE function along the UE-dimension.


\begin{remark}\emph{
	The WMMSE algorithm is only one of the sub-optimal algorithms for solving the SE-maximization precoding problem in MU-MISO systems. One may wonder \emph{whether other algorithms for optimizing  problem \textsf{P-S} have the same structure?} To answer the question, we have proved that other two algorithms for solving problem \textsf{P-S}, fractional programming (FP) and GD algorithms, can also be re-expressed as the two recursions in Proposition \ref{prop:reexp-miso-prec}. The difference of their RIEs from the WMMSE algorithm only lies in the elements in $\mathbf{d}_k^{(\ell)}$. One may also wonder \emph{whether the algorithms for optimizing other precoding problems (with different objective functions and constraints) have this structure}?
	To answer this question, we have proved that the RIE of a FP algorithm for solving the precoding problem that maximizes the minimum data rate of UEs is the same as Proposition \ref{prop:reexp-miso-prec}, and the difference only lies in the elements in $\mathbf{d}_k^{(\ell)}$. The proofs are not provided due to the limited space. This implies that the structure of an RIE
depends on the type and relation of sets in a problem.}
\end{remark}



\textbf{Example 2 (Precoding for MU-MIMO system)}: We extend the precoding problem into a MU-MIMO system, where each UE is equipped with $N_{\sf U}$ antennas and receives $M$ data streams. The precoding matrix can be optimized, say from the following SE-maximization problem \cite{WMMSE2011Shi},
\begin{subequations}\label{eq:prob}
	\begin{align}
		\hspace{-2.5mm}\textsf{P-M}:& \max_{\mathbf{W}_1,\cdots,\mathbf{W}_K}  \textstyle\sum_{k=1}^K \log_2\det \big(\mathbf{I}_{N_{\sf U}}+\mathbf{H}_k\mathbf{W}_{k}\mathbf{W}_{k}^H\mathbf{H}_k^H\cdot\notag\\
		& \vspace{15mm} \big(\textstyle\sum_{j=1,j\neq k}^K\mathbf{H}_k\mathbf{W}_{j}\mathbf{W}_{j}^H\mathbf{H}_k^H + \sigma_0^2\mathbf{I}_{N_{\sf U}} \big)^{-1}\big) \label{eq:prob-MIMO} \\
	&	{\mathrm{s.t.}} ~
		  \textstyle\sum_{k=1}^K\|\mathbf{W}_k \|^2_F \leq P_{\max},
	\end{align}
\end{subequations}
where  $\mathbf{W}_{k}\in\mathbb{C}^{N_{\sf B}\times M}$ and $\mathbf{H}_k\in\mathbb{C}^{N_{\sf U}\times N_{\sf B}}$ are the precoding matrix and channel matrix of UE$_k$, respectively. The SE can be achieved by the minimum-mean-squared-error combiner at each UE \cite{WMMSE2011Shi}.

Denote the precoding policy as $\mathbf{W}^{\star} = \phi_\mathsf{M}(\mathbf{H})$, where $\mathbf{W}^{\star}=[\mathbf{W}_{1}^{\star},\cdots,\mathbf{W}_{K}^{\star}] \in\mathbb{C}^{N_{\sf B}\times M K}$ is the optimized precoding matrix, and $\mathbf{H}\triangleq[\mathbf{H}_1^T,\cdots,\mathbf{H}_K^T]^T \in\mathbb{C}^{K N_{\sf U}\times N_{\sf B}}$. As shown in \eqref{eq:prob-MIMO}, both MUI and IDI are reflected in $\mathbf{H}_k\mathbf{W}_{j}\mathbf{W}_{j}^H\mathbf{H}_k^H=\sum_{p=1}^M\mathbf{H}_k\mathbf{w}_{p_j}\mathbf{w}_{p_j}^H\mathbf{H}_k^H$ instead of reflecting in the environmental
parameters (i.e., $\mathbf{H}$) of the policy.

There are three sets in the problem: antennas at the BS (AN$^{\tt BS}$s for short), data streams (DSs) and antennas at UEs (AN$^{\tt UE}$s for short), since $\mathbf{W}^{\star}$ is with AN$^{\tt BS}$- and DS-dimensions, meanwhile $\mathbf{H}$ is with AN$^{\tt BS}$- and
AN$^{\tt UE}$-dimensions \cite{LSJ_MultiDim_GNN_2022}. Their types and relations are as follows.
\begin{itemize}
	\item \emph{Type of sets}: The AN$^{\tt BS}$ set is a normal set. The other two sets are nested sets each consisting of $K$ subsets, i.e., \\
	DS set: \{\!\{DSs of UE$_1$\},...,\{DSs of  UE$_K$\}\!\}, \\
	AN$^{\tt UE}$ set: \{\!\{ANs at UE$_1$\},...,\{ANs at UE$_K$\}\!\}.
	\item \emph{Relation among sets}: The subsets in the DS set and AN$^{\tt UE}$ set can only be permuted in the same way.
\end{itemize}
Consequently, the precoding policy $\phi_\mathsf{M}(\cdot)$ satisfies a \emph{nested and joint permutation property} as follows,
\begin{equation}\label{eq:prec-pe}
	\mathbf{\Pi}_{\sf B}^T\mathbf{V^{\star}}\mathbf{\Omega}_{\sf D}=\phi_\mathsf{M}(\mathbf{\Omega}_{\sf U}^T\mathbf{H}\mathbf{\Pi}_{\sf B}),
\end{equation}
where $\mathbf{\Pi}_{\sf B}$ is the permuting matrix that changes the order of AN$^{\tt BS}$s, $\mathbf{\Omega}_{\sf D}\triangleq(\mathbf{\Pi}_{\sf UE}\otimes \mathbf{I}_M)\mathsf{diag}(\mathbf{\Pi}_{\mathsf{D},1}, \cdots, \mathbf{\Pi}_{\mathsf{D},K})$ changes the order of UEs with $\mathbf{\Pi}_{\sf UE}$ and changes the order of DSs of UE$_k$ with $\mathbf{\Pi}_{\mathsf{D},k}$, $\mathbf{\Omega}_{\sf U}\triangleq(\mathbf{\Pi}_{\sf UE}\otimes \mathbf{I}_M)\mathsf{diag}(\mathbf{\Pi}_{\mathsf{U},1}, \cdots, \mathbf{\Pi}_{\mathsf{U},K})$ changes the order of UEs with $\mathbf{\Pi}_{\sf UE}$ and changes the order of ANs at UE$_k$ with $\mathbf{\Pi}_{\mathsf{U},k}$.

%


Problem \textsf{P-M} is a non-convex problem and a sub-optimal solution can also be obtained via the WMMSE algorithm \cite{WMMSE2011Shi}. Under the assumption that the signal-to-noise ratio (SNR) is high and by approximating the matrix inverse with its Taylor's expansion, the iterative equations of the algorithm in the $(\ell+1)$-th iteration can be approximated as \cite{GJ-MLSP},
\begin{subequations}
	\begin{align}
		\!\!\!\!\mathbf{u}_{m_k}^{(\ell+1)} \!&\approx\! 2\mathbf{H}_k\mathbf{w}_{m_k}^{(\ell)} -\textstyle\sum_{p=1,p\neq m}^M (\mathbf{w}_{p_k}^{H(\ell)}\mathbf{H}_k^H\mathbf{H}_k\mathbf{w}_{m_k}^{(\ell)}\!)\mathbf{H}_k\mathbf{w}_{p_k}^{(\ell)} \notag\\
		&\!-\!\!\! \textstyle\sum_{j=1,j\neq k}^K \textstyle\sum_{p=1}^M (\mathbf{w}_{p_j}^{H(\ell)}\mathbf{H}_k^H\mathbf{H}_k\mathbf{w}_{m_k}^{(\ell)})\mathbf{H}_k\mathbf{w}_{p_j}^{(\ell)}\!,\label{eq:upd-u-2}\\
		\!\!\!\!\mathbf{w}_{m_k}^{(\ell+1)} &\approx 2\mathbf{H}_k^H\mathbf{u}_{m_k}^{(\ell)} - \textstyle\sum_{p=1,p\neq m}^M (\mathbf{u}_{p_k}^{H(\ell)}\mathbf{H}_j\mathbf{H}_k^H\mathbf{u}_{m_k}^{(\ell)})\mathbf{H}_k^H\mathbf{u}_{p_k}^{(\ell)} \notag\\
		&\!- \textstyle\sum_{j=1,j\neq k}^K\sum_{p=1}^M (\mathbf{u}_{p_j}^{H(\ell)}\mathbf{H}_j\mathbf{H}_k^H\mathbf{u}_{m_k}^{(\ell)})\mathbf{H}_j^H\mathbf{u}_{p_j}^{(\ell)}, \label{eq:upd-v-2}
	\end{align}
\end{subequations}
where $\mathbf{u}_{m_k}^{(\ell)}$ and $\mathbf{w}_{m_k}^{(\ell)}$ are respectively the updated combining and precoding vectors for the $m$-th DS at UE$_k$.

In \eqref{eq:upd-u-2} and \eqref{eq:upd-v-2}, the inner products in the two summation terms respectively reflect the IDI of the UE$_k$ and the IDI of all the UEs (reflecting MUI implicitly).

After omitting the
representation dimension, the updated variables in the $\ell$-th iteration and the channel coefficient can be expressed as a tensor $\vec{\mathbf{D}}^{(\ell)}=[\mathbf{D}_{1_1}^{(\ell)},\cdots,\mathbf{D}_{M_K}^{(\ell)}]\in{\mathbb C}^{N_{\mathsf{B}}\times KN_{\mathsf{U}}\times KM}$ with AN$^{\tt BS}$-,
AN$^{\tt UE}$-, and DS-dimensions, where each element in representation matrix $\mathbf{D}_{m_k}^{(\ell)}$ is,
\begin{equation}
	d_{n, m_k, e_j}^{(\ell)} =
	\begin{cases}
		\mathsf{concat}(u_{e_k, m_k}^{(\ell)}, w_{n,m_k}^{(\ell)}, h_{e_k, n}) & j=k, \\
		0 & j\neq k,
	\end{cases} \notag
\end{equation}
$u_{e_k, m_k}^{(\ell)}$ is the $e$-th element in $\mathbf{u}_{m_k}^{(\ell)}$, $w_{n,m_k}^{(\ell)}$ is the $n$-th element in $\mathbf{w}_{m_k}^{(\ell)}$, and $ h_{e_k, n}$ is the channel coefficient from the $n$-th AN at the BS to the $e$-th AN at UE$_k$.

\begin{proposition}\label{prop:reexp-mimo-prec}
	\emph{ The iterative equations in \eqref{eq:upd-u-2} and \eqref{eq:upd-v-2} can be re-expressed as $\vec{\mathbf{D}}^{(\ell+1)}=\psi_{\sf M}(\vec{\mathbf{D}}^{(\ell)})\odot \mathbf{I}_{\sf B}+\vec{\mathbf{D}}^{(\ell)}\odot(1-\mathbf{I}_{\sf B})$, where $\mathbf{I}_{\sf B}\triangleq\mathsf{diag}(\mathbf{1}_{N_{\mathsf{UE}}\times M},\cdots,\mathbf{1}_{N_{\mathsf{UE}}\times M})$, and $\psi_{\sf M}(\vec{\mathbf{D}}^{(\ell)})$ can be expressed in the form of three recursions as follows. \\
		\underline{\emph{First recursion}}: $\vec{\mathbf{B}}^{(\ell+1)}=\psi_{\sf M}(\vec{\mathbf{D}}^{(\ell)})$ is an NPE function-II along the DS-dimension as,
		\begin{align}
			\mathbf{B}_{m_k}^{(\ell+1)} &= \textstyle f_\mathsf{M}\Big(\mathbf{D}_{m_k}^{(\ell)},\sum_{p=1,p\neq m}^M q_\mathsf{M,1}\big(\mathbf{D}_{m_k}^{(\ell)}, \mathbf{D}_{p_k}^{(\ell)}\big), \notag\\
			&\hspace{10mm}\textstyle\sum_{j=1,j\neq k}^K\sum_{p=1}^M q_\mathsf{M,2}\big(\mathbf{D}_{m_k}^{(\ell)}, \mathbf{D}_{p_j}^{(\ell)}\big)\Big), \notag
		\end{align}
        where $\vec{\mathbf{B}}^{(\ell+1)}=[\mathbf{B}_{1_1}^{(\ell)},\cdots,\mathbf{B}_{M_K}^{(\ell)}]\in{\mathbb C}^{N_{\mathsf{B}}\times KN_{\mathsf{U}}\times KM}$, and the combiner
        $f_{\sf M}(\cdot)$ and processors $q_{\sf M,1}(\cdot), q_{\sf M,2}(\cdot)$ are DS-size invariant functions.\\
		\underline{\emph{Second recursion}}: $f_\mathsf{M}(\mathbf{X})$, $q_{\mathsf{M},1}(\mathbf{X})$ and $q_{\mathsf{M},2}(\mathbf{X})$ are NPE functions-I along the AN$^{\tt UE}$-dimension, where the processors and combiners are DS- and AN$^{\tt UE}$-size invariant.\\
		\underline{\emph{Third recursion}}:
		The processors and combiners in the NPE functions-I of the second recursion are APE functions-I along the AN$^{\tt BS}$-dimension, where the processors and combiners are DS-, AN$^{\tt UE}$-, AN$^{\tt BS}$-size invariant.
	}
	\begin{IEEEproof}
		Omitted due to the similarity to Appendix \ref{proof:prop:reexp-power-bandw-allo}.
	\end{IEEEproof}
\end{proposition}

The NPE functions-II/I in the first and the second recursions only have the following difference from \eqref{eq:nest-pe-I} and \eqref{eq:nest-pe-II}: $q_{\rm I, 2}(\cdot)$ and $q_{\rm II, 2}(\cdot)$ in \eqref{eq:nest-pe-I} and \eqref{eq:nest-pe-II} become linear (in particular, $q_{\rm I, 2}(\mathbf{x})=\mathbf{x},q_{\rm II, 2}(\mathbf{X})=\mathbf{X}$) in the second and the first recursions.

The function $\phi_{\sf M}(\cdot)$ satisfies the permutability to the DSs, AN$^{\tt UE}$s and AN$^{\tt BS}$s separately in the three recursions. The joint permutability of subsets in the DS set and AN$^{\tt UE}$ set is satisfied by the output function $\vec{\mathbf{D}}^{(\ell+1)}=\psi_{\sf M}(\vec{\mathbf{D}}^{(\ell)})\odot \mathbf{I}_{\sf B}+\vec{\mathbf{D}}^{(\ell)}\odot(1-\mathbf{I}_{\sf B}) \triangleq f_{\rm out}(\psi_{\sf M}(\vec{\mathbf{D}}^{(\ell)}),\vec{\mathbf{D}}^{(\ell)})$.

\begin{figure}[!htb]
	\centering
	\includegraphics[width=0.35\linewidth]{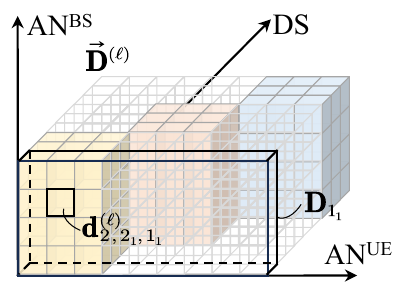}
	\vspace{-1mm}
	\caption{An example of $\vec{\mathbf{D}}^{(\ell)}$ for $N_{\sf B}=4, K=3, M=3, N_{\sf U}=3$. Each cube is an element in the updated tensor, say ${d}_{n,m_k,e_k}^{(\ell)}$. The cube without color is zero.} 
	\label{fig:3d-rep}
	\vspace{-0.1mm}
\end{figure}

To help understand the RIE, the tensor $\vec{\mathbf{D}}^{(\ell)}$ is illustrated in Fig. \ref{fig:3d-rep}. We can see that only the diagonal blocks in the tensor $\vec{\mathbf{D}}^{(\ell)}$ are updated, while the non-diagonal blocks are zero.

\vspace{-0.2mm}

\begin{figure}[!htb]
	\centering
	\includegraphics[width=.95\linewidth]{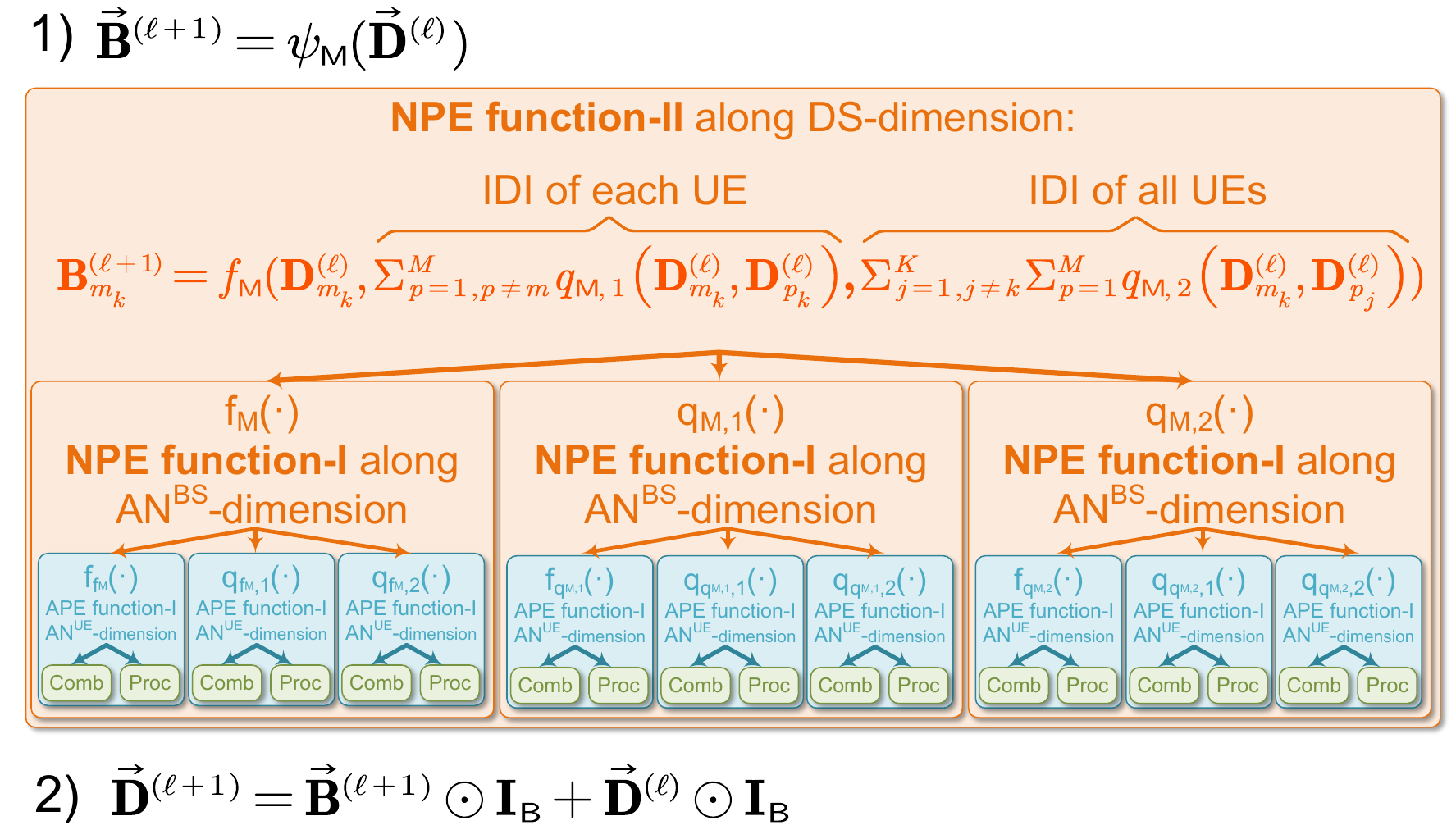}\vspace{-2mm}
	\caption{Structure of the RIE of the WMMSE algorithm for problem \textsf{P-M}, where ``Comb'' and ``Proc'' stand for combiner and processor.}
	\vspace{-0.2mm}
	\label{fig:rie-pm}
\end{figure}

The structure of the RIE is shown in Fig. \ref{fig:rie-pm}. Again, we can
see how the I-O relation of each iteration is composed of one-set PE functions recursively and an output function.

\vspace{-0.2mm}

\vspace{1mm}\textbf{Observation 3}. In the RIEs in Proposition \ref{prop:reexp-mimo-prec}, the attention processor is only in the NPE function-II along the DS-dimension. In the first recursion, $q_{\mathsf{M},1}(\cdot)$ reflects the IDI of each UE, and $q_{\mathsf{M},2}(\cdot)$ reflects the IDI of all UEs (i.e., MUI).

\vspace{-1mm}\begin{remark}\emph{
In addition to MUI and IDI, there may exist other types of interference, say inter-cell interference (ICI) in a multi-cell MU-MISO system. In this system, the UEs constitute a nested set as \{\{UEs in Cell$_1$\},...,\{UEs in Cell$_{N_{\sf C}}$\}\}, where Cell$_{n}$ denotes the $n$-th cell. By re-expressing the iterative equation of an algorithm for a precoding problem (say WMMSE for maximizing SE), we find that an NPE function-II along UE-dimension is in the first recursion, where the attention processor captures both the ICI and MUI.}
\end{remark}\vspace{-1mm}

From Observations 2 and 3 and Remark 4, we can infer that in the RIE of an algorithm for optimizing a policy, when the interference is not reflected in the environmental parameters, \emph{attention appears along the dimension with interference}.

\vspace{1mm}\subsubsection{\underline{Interference is reflected in environmental parameters}}

~

We take optimizing power control in an interference system with $K$ transmitters (Txs) and $K$ receivers (Rxs) as an example. The transmit powers at the Txs can be optimized, say from the following problem,
\begin{subequations}\label{P: max sum-rate1}
	\begin{align}
		\textsf{P-C}: \max_{p_1,\cdots,p_K} & \textstyle\sum_{k=1}^K \log_2\Big(1+\frac{|g_{kk}|^2 p_k}{\sum_{j=1,j\neq k}^K |g_{kj}|^2 p_j + \sigma_0^2}\Big)\label{P:object}\\
		{\rm s.t.} ~~& 0\leq p_k\leq P_{\max}, k=1,\cdots,K, \label{P: constraint}
	\end{align}
\end{subequations}
where $g_{kj}$ is the channel coefficient from the Tx$_k$ to the Rx$_j$.

Denote the power control policy as $\mathbf{p}^\star=\phi_\mathsf{C}(\mathbf{G})$, where $\mathbf{p}^\star=[p_1^\star,\cdots,p_K^\star]^T$, and $\mathbf{G}=[|g_{kj}|]\in{\mathbb C}^{K\times K}$.

The sets in the problem and their relation are as follows.
\begin{itemize}
	\item \emph{Type of sets}: Tx set and Rx set. Both are normal sets.
	\item \emph{Relation among sets}: The elements in the two sets need to be jointly permuted.
\end{itemize}
As a result, the policy satisfies the following joint PE property, ${\bm\Pi}^T\mathbf{p}^\star=\phi_\mathsf{C}({\bm\Pi}^T\mathbf{G}{\bm\Pi})$ \cite{Eisen2020,SYF}.

Different from the MU-MISO or MU-MIMO precoding policies (i.e., $\phi_\mathsf{S}(\cdot)$ and $\phi_\mathsf{M}(\cdot)$), the interference is reflected in the environmental parameters of the power control policy, i.e., $\mathbf{G}$. Specifically, $g_{kj}, k\neq j$ is the interference channel coefficient from the $k$-th transmitter to the $j$-th receiver.

Problem \textsf{P-C} is non-convex and a sub-optimal solution can also be obtained via the WMMSE algorithm. The updated variables in the $\ell$-th iteration and the channel matrix can be expressed as a matrix $\mathbf{D}^{(\ell)}=[\mathbf{d}_1^{(\ell)},\cdots,\mathbf{d}_K^{(\ell)}]\in{\mathbb R}^{K\times K}$, where each element is,
\begin{equation}\label{eq:upd-var}
	d_{jk}^{(\ell)} = \begin{cases}
		\mathsf{concat}(v_k^{(\ell)}, u_k^{(\ell)}, w_k^{(\ell)},|g_{kk}|) & j=k, \\
		\mathsf{concat}(|g_{jk}|, |g_{kj}|) & j\neq k,
	\end{cases}
\end{equation}
$v_k^{(\ell)},u_k^{(\ell)},w_k^{(\ell)}$ are intermediate variables updated in the $\ell$-th iteration of the algorithm. The optimized power in the $\ell$-th iteration is $p_k^{(\ell)}=|v_k^{(\ell)}|^2$.


\begin{proposition}\label{prop:reexp-pc}\emph{
		The iterative equations of the WMMSE algorithm for problem \textsf{P-C} can be re-expressed as $\mathbf{D}^{(\ell+1)}=\psi_{\sf C}(\mathbf{D}^{(\ell)})\odot\mathbf{I}_K + \mathbf{D}^{(\ell)}\odot (1-\mathbf{I}_K)$, where $\psi_{\sf C}(\mathbf{D}^{(\ell)})$ can be expressed in the form of two recursions as follows.\\
		\underline{\emph{First recursion}}: $\mathbf{B}^{(\ell+1)}=\psi_{\sf C}(\mathbf{D}^{(\ell)})$ is an APE function-I along the Tx-dimension as,
		\begin{equation}\label{eq:upd-pc}
			\mathbf{b}_k^{(\ell+1)} = f_{\mathsf{C}}\big(\mathbf{d}_k^{(\ell)}, \textstyle\sum_{j=1}^K q_{\mathsf{C}}(\mathbf{d}_j^{(\ell)})\big),
		\end{equation}
        where $\mathbf{B}^{(\ell+1)}=[\mathbf{b}_1^{(\ell+1)},\cdots,\mathbf{b}_K^{(\ell+1)}]$,
        $f_{\sf C}(\cdot)$ and $q_{\sf C}(\cdot)$ are Tx-size invariant functions.\\
		\underline{\emph{Second recursion}}:
		$\mathbf{y}= q_{\mathsf{C}}(\mathbf{d}_k^{(\ell)})$ and $\mathbf{y}=f_{\mathsf{C}}(\mathbf{x})$ are APE functions-I along the Rx-dimension, where the combiners and processors are Tx- and Rx-size invariant. \\
	}\vspace{-3mm}
	\begin{IEEEproof}
		Omitted due to the similarity to Appendix \ref{proof:prop:reexp-power-bandw-allo}.
	\end{IEEEproof}
\end{proposition}

%

Similar to the RIE of the WMMSE algorithm for MU-MIMO precoding, the function $\psi_{\sf C}(\cdot)$ satisfies the permutability to the Txs and Rxs separately in the two recursions, while the joint PE property is satisfied by the output function $\mathbf{D}^{(\ell+1)}=\psi_{\sf C}(\mathbf{D}^{(\ell)})\cdot\mathbf{I}_K + \mathbf{D}^{(\ell)}\cdot(1-\mathbf{I}_K) \triangleq f_{\rm out}(\psi_{\sf C}({\mathbf{D}}^{(\ell)}),{\mathbf{D}}^{(\ell)})$.


\vspace{1mm}\textbf{Observation 4}. The attention processor does not appear in both recursions of the RIE.
\vspace{1mm}

To explain why attention does not appear for reflecting the interference in this system, we regard problem \textsf{P-C} as a degenerated version of the precoding problem \textsf{P-S} in MU-MISO systems, where a BS with $N_{\sf B}$ antennas serves $K$ single-antenna UEs. This comes from the fact that the $K\times K$ square matrix $\mathbf{G}$ can always be decomposed into the product of two matrices as $\mathbf{G}=\mathbf{W}^H\mathbf{H}$, where $\mathbf{W}=[\mathbf{w}_1,\cdots,\mathbf{w}_K]$, $\|\mathbf{w}_k\|=1$, and $\mathbf{H}=[\mathbf{h}_1,\cdots,\mathbf{h}_K]$ are $N_{\sf B}\times K$ matrices, as long as $N_{\sf B}\geq K$ \cite{matrix-theory}. Then, $g_{kj}=\mathbf{w}_j^H\mathbf{h}_k$,
 where $\mathbf{w}_k$ and $\mathbf{h}_j$ can be respectively seen as the beamforming vector for UE$_k$ and the channel vector of UE$_j$. In the degenerated problem, only the transmit powers for the UEs need to be optimized while the beamforming matrix is pre-determined. The iterative equations of the WMMSE algorithm for optimizing the transmit powers can be expressed as \cite{WMMSE2011Shi},
\begin{subequations}\label{eq:vuw-1}
	\begin{align}
		v_k^{(\ell+1)} &= \left[\frac{z_k^{(\ell)}u_k^{(\ell)}|\mathbf{w}_k^H\mathbf{h}_k|}{\sum_{j=1}^K z_j^{(\ell)}(u_j^{(\ell)})^2 |\mathbf{w}_k^T\mathbf{h}_j|^2}\right]_0^{\sqrt{P_{\max}}}, \label{eq:v-upd-1}\\
		u_k^{(\ell+1)} &= \frac{v_k^{(\ell)}|\mathbf{w}_k^H\mathbf{h}_k|}{\sum_{j=1}^K (v_j^{(\ell)}|\mathbf{w}_j^H\mathbf{h}_k|)^2 + \sigma_0^2}, \label{eq:u-upd-1}\\
		z_k^{(\ell+1)} &= \frac{1}{1-u_k^{(\ell)}v_k^{(\ell)}|\mathbf{w}_k^H\mathbf{h}_k|}. \label{eq:w-upd-1}
	\end{align}
\end{subequations}
By denoting $\mathbf{d}_k^{(\ell)}=\mathsf{concat}(\mathbf{w}_k,v_k^{(\ell)},u_k^{(\ell)},z_k^{(\ell)},\mathbf{h}_k)$ as representation vector, 
the RIE of \eqref{eq:vuw-1} can be obtained as,
\begin{equation}\label{degerATE}
	\mathbf{d}_k^{(\ell+1)} = f_{\mathsf{S'}}\Big(\mathbf{d}_k^{(\ell)},\textstyle\sum_{j=1,j\neq k}^K q_{\mathsf{S'}}(\mathbf{d}_k^{(\ell)},\mathbf{d}_j^{(\ell)})\Big),
\end{equation}
where $q_{\mathsf{S'}}(\mathbf{d}_k^{(\ell)},\mathbf{d}_j^{(\ell)}) = [z_j^{(\ell)}(u_j^{(\ell)})^2 |\mathbf{w}_k^H\mathbf{h}_j|^2, (v_j^{(\ell)}|\mathbf{w}_j^H\mathbf{h}_k|)^2]^T$.

If the environment parameter is $\mathbf{H}$, then the RIE in \eqref{degerATE} is with an attention processor. However, since the environment parameter for the power allocation problem is  $\mathbf{G}$ that consists of
the interference coefficients $g_{kj}=\mathbf{w}_j^H\mathbf{h}_k, k \neq j$, $q_{\mathsf{S'}}(\mathbf{d}_k^{(\ell)},\mathbf{d}_j^{(\ell)})$ in \eqref{eq:upd-pc} is not an attention processor.

\vspace{-2mm}
\subsection{Summary of the Analyses}\label{sec:summary}\vspace{-1mm}
Previous analyses demonstrate that the iterative equations of numerical algorithms for solving wireless problems with sets can be re-expressed as the RIEs.
The structure of an RIE is determined by the sets, and does not depend on specific algorithms and specific objective function/constraints. 

Moreover, \emph{the I-O relation of an RIE after convergence is the I-O relation of a policy with permutation property}. This means that the RIE captures the inherent structure of the policy. The characteristics of the RIE structure are summarized as follows.

\emph{(1) When does the attention processor appear?}

\begin{itemize}
	\item For the problem in interference-free scenarios, the attention processor is not in the RIE.
	\item For the problems in interference scenarios,
	\begin{itemize}
		\item If interference is reflected in the environmental parameters, then the attention processor is not in the RIE.
		\item Otherwise, the attention processor is in the RIE along the dimensions with interference.
	\end{itemize}
\end{itemize}

\emph{(2) How is the order of recursions determined?}
When the interference is not reflected in the environmental parameters, the first recursion of each RIE corresponds to the dimension with interference. The interference is reflected by the attention processor, as exemplified by the two precoding problems.
\begin{itemize}
    \item For problem \textsf{P-S}, the MUI power is $|\mathbf{h}_k^{H}\mathbf{w}_{i}|^2$. In Proposition \ref{prop:reexp-miso-prec}, the processor in the first recursion that computes the attention between two vectors can capture MUI. If the attention is computed in the second recursion, then the attention between two elements (instead of two vectors) is computed, which cannot reflect the MUI.
    \item For problem \textsf{P-M}, the MUI and IDI are reflected in $\mathbf{H}_k\mathbf{W}_{j}\mathbf{W}_{j}^H\mathbf{H}_k^H$. In Proposition \ref{prop:reexp-mimo-prec}, the processor in the first recursion that computes the attention between two matrices can reflect the MUI and IDI. The interference cannot be reflected if the attention is computed in the second or the third recursions.
\end{itemize}

Except for the first recursion, the order of other recursions does not matter, because the interference is already reflected in the first recursion. For example, the iterative equations for problem \textsf{P-M} can also be re-expressed in another form, where the one-set PE functions in the second and third recursions are respectively along the AN$^{\tt BS}$- and AN$^{\tt UE}$-dimensions.

\emph{(3) Processors and combiners are invariant to the size of which dimensions?}
In each recursion, the combiners and processors in the one-set PE functions are invariant not only to the size of the corresponding dimension, but also invariant to the sizes of all the dimensions in previous recursions.
Hence,
the combiners and processors in the last recursion are invariant to the sizes of all the dimensions. Since the I-O relation of an RIE can be expressed as a composite function of the combiner and processor in the last recursion as shown in
Propositions \ref{prop:reexp-power-bandw-allo}, \ref{prop:reexp-miso-prec}, \ref{prop:reexp-mimo-prec}, and \ref{prop:reexp-pc},
the RIE is  ``size generalizable''.

%

%
%

\vspace{-1mm}\begin{remark}\emph{
Such a structure of the RIE (and hence the policy with permutation property) is identified only from several algorithms of several problems. Nonetheless, the approach of revealing the structure of the policy can be applied to other algorithms, other problems, and various settings, which differs from the structures of optimal solutions adopted in \cite{Bipartie_GNN_BF_TWC2023,li2024hpe}. }\end{remark}

\vspace{-2mm}
\section{Efficient GNNs with Attention Mechanism for Learning Wireless Policies}\label{sec:gnn-design-framework}\vspace{-1mm}

In this section, we propose a framework to incorporate the attention mechanism into GNNs for learning wireless policies efficiently. We start by showing that the I-O relations of the update equations of GNNs and a Transformer are one-set PE functions. We proceed to provide the framework for the policies inspired by their structures,
and demonstrate how to use the framework with a concrete example.

	\vspace{-3mm}
	\subsection{GNNs and the Transformer: Parameterized PE Functions}\vspace{-0.1mm}
For a GNN that updates the hidden representations of vertices (or edges) in each layer, the I-O relation of the update equation can be designed as a \emph{parameterized} PE function, where the combiner(s) and processor(s) can be parameterized functions.

For example, the update equation of a graph convolutional neural network (GCN) learning over a graph with $K$ vertices can be expressed as,
	\begin{align}\label{eq:upd-vgcn}
		\bar{\mathbf{d}}_k^{(\ell+1)} = \sigma(\mathbf{V}\bar{\mathbf{d}}_k^{(\ell)}+\textstyle\sum_{j=1,j\neq k}^K \mathbf{U}\bar{\mathbf{d}}_j^{(\ell)}),
	\end{align}
	where $\bar{\mathbf{d}}_k^{(\ell)}$ is the updated representation of the $k$-th vertex in the $\ell$-th layer, $\sigma(\cdot)$ is an activation function, $\mathbf{V}$ and $\mathbf{U}$ are weight matrices in the combiner and the linear processor. The summation $\sum(\cdot)$ is the \emph{pooling function}.

We can see that the I-O relation of the update equation is a parameterized APE function-I in \eqref{eq:1d-pe-I}, where $f_{\rm I}(\cdot)$ is a linear function cascaded by an activation function, and $q_{\rm I}(\cdot)$ is a linear ordinary processor.

	The encoder-only Transformer without positional encoding can be regarded as a GNN learning over a homogeneous complete graph \cite{gradient}.
Specifically, the representation of the $k$-th token in the $(\ell+1)$-th layer is updated as follows,
	\begin{align}
		\bar{\mathbf{d}}_k^{(\ell+1)}\!\! &= \mathsf{FFN}\Big(\bar{\mathbf{d}}_k^{(\ell)}+\xi\big((\mathbf{U}_{\mathsf{q}}\bar{\mathbf{d}}_k^{(\ell)})^T(\mathbf{U}_{\mathsf{k}}\bar{\mathbf{d}}_k^{(\ell)})\big)(\mathbf{U}_{\mathsf{v}}\bar{\mathbf{d}}_k^{(\ell)})\notag\\
		&+\!\!\textstyle\sum_{j=1,\! j\neq k}^K\!\xi\big(\!(\mathbf{U}_{\mathsf{q}}\bar{\mathbf{d}}_k^{(\ell)}\!)^T\!(\mathbf{U}_{\mathsf{k}}\bar{\mathbf{d}}_j^{(\ell)}\!)\!\big)(\mathbf{U}_{\mathsf{v}}\bar{\mathbf{d}}_j^{(\ell)}\!)\!\Big)\notag\\
		&=\mathsf{FFN}\big(\bar{\mathbf{d}}_k^{(\ell)},\textstyle\sum_{j=1,j\neq k}^K Q(\bar{\mathbf{d}}_k^{(\ell)},\bar{\mathbf{d}}_j^{(\ell)})\big),\label{eq:transformer-upd}
	\end{align}
	where $\mathsf{FFN}(\cdot)$ is a FNN, $\xi(\cdot)$ is the softmax function \cite{Transformer}, $\mathbf{U}_{\mathsf{q}}, \mathbf{U}_{\mathsf{k}}$ and $\mathbf{U}_{\mathsf{v}}$ are weight matrices.

We can see that the I-O relation of the update equation in \eqref{eq:transformer-upd} is a parameterized APE function-II in \eqref{eq:1d-pe-II}, where the combiner $f_{\rm II}(\cdot)$ is a FNN, and $q_{\rm II}(\cdot)=Q(\bar{\mathbf{d}}_k^{(\ell)},\bar{\mathbf{d}}_j^{(\ell)})$ is an attention processor.

\vspace{-1mm}\begin{remark}\emph{
The parameterized attention processor in the Transformer can approximate an ordinary processor. Hence, if a policy on one set can be well-learned by a GCN, then the policy can also be  well-learned by the Transformer, but the converse statement may not be true. However, the Transformer only satisfies the APE property, hence it is inefficient for learning a multitude of wireless policies on a nested set or on more than one set. }\end{remark}

	
	

\vspace{-3mm}
\subsection{A Framework of Designing GNNs}\vspace{-0.1mm}	

Recall that the RIE of an algorithm for each wireless problem in section \ref{sec:design-gnns} consists of one-set PE functions. On the other hand, the I-O relation of the update equation of GNNs
can be designed as a parameterized PE function. This suggests that a GNN for learning a wireless policy from a problem can be designed to ``align with'' the RIE.

Specifically, after re-expressing the iterative equations of an algorithm as an RIE, the update equation of the GNN can be designed with a recursive structure, where the types of one-set PE functions in each recursion are the same as the RIEs. To make the GNN trainable and size generalizable, the processors and combiners in the last recursion (that are invariant to the sizes of all the dimensions) are designed as FNNs.
Since RIE captures the essential structure of a policy with PE property, such a framework is referred to as ``\emph{PE structure alignment}''.

However, for many problems, it might be tedious to design or find the numerical algorithms for solving the problems, or to re-express the iterative equations of the  algorithms. Fortunately, a GNN can be designed as long as the structures of RIEs can be determined, without the need to design the algorithms and derive their RIEs. This indicates that an efficient GNN can be designed directly from the problem.

From the analyses in section
\ref{sec:design-gnns}, we can see that the structure of an RIE can be determined by analyzing the characteristics of an optimization problem
with the following steps.
\begin{enumerate}[leftmargin=.4cm]
	\item Identify all the sets in the problem by using the method proposed in \cite{LSJ_MultiDim_GNN_2022}. This determines the number of recursions in the RIE.
	\item Identify the type of each set (i.e., normal set or nested set), and the relation among sets (i.e., independent or joint) by using the method in \cite{LSJ_MultiDim_GNN_2022}. This determines the type of one-set PE function in each recursion, and whether an output function is required.
	\item Identify whether there exists interference in the problem, and whether it is reflected in the environmental parameters. This determines whether an attention processor is used.
\end{enumerate}


Then, the GNN can be designed to align with the structure of the RIE. The framework of designing a
GNN from an optimization problem is summarized in Procedure \ref{proc:design-dnn}.

\begin{algorithm}
	\caption{Designing the update equation of efficient GNN}\label{proc:design-dnn}
	{\footnotesize
		\begin{algorithmic}[1]
			\State Formulate an optimization problem and define the resulting wireless policy
			\State Identify the number of sets $S$ in the problem, the type of each set and the relation among sets
			\Statex \emph{Design update equation of GNN for updating $\vec{\mathbf{D}}^{(\ell)}$:}
			\For {$s=1:S$}
			\State Select one set from all the sets. If there exists interference in the $p$-th dimension and the interference is not reflected in the environmental parameters of the problem, then the $p$-th set is selected in the first recursion.
			\If {the selected set is a normal set}
			\State Design the one-set PE function(s) in the $s$-th recursion as APE \Statex \hspace{7mm} function(s)
			\Else {~the selected set is a nested set}
			\State Design the one-set PE function(s) in the $s$-th recursion as NPE \Statex \hspace{7mm} function(s)
			\EndIf
			\If {interference exists in the selected dimension, and the interference is not reflected in the environmental parameters of the problem}
			\State Design the processor(s) as attention processor(s)
			\Else
			\State Design the processor(s) as ordinary processor(s)
			\EndIf
            \If {$s<S$}
			\State Design combiner(s) and processor(s) as one-set PE functions
            \Else
            \State Design combiner(s) and processor(s)  as FNNs
            \EndIf
			\EndFor
			\If {the elements in the $s_1,\cdots,s_p$-th sets should be jointly permuted}
			\State only the elements in $\vec{\bf D}^{(\ell)}$ with indices satisfying $n_{s_1}=\cdots=n_{s_p}$ \Statex \hspace{3mm} are updated
			\EndIf	
	\end{algorithmic}}
\end{algorithm}
\vspace{-0.1mm}

One may ask what happens if the update equation of the GNN is not aligned with the RIE. If the number of recursions of the update equation or the type of one-set PE functions are not the same as those of the RIE, then the permutation property satisfied by the GNN does not match with the property of the policy, which reduces the learning efficiency or may degrade the learning performance. For each one-set PE function in each recursion, if the processor in the GNN is not the same as the processor in the RIE, the learning performance and efficiency will be degraded or the computational complexity of the GNN will be high. For instance,
we can simply design the processors in the one-set PE functions of all the recursions as parameterized attention processors as
in \cite{GJ-TMLCN}, which can approximate ordinary processors.
However, such a design incurs higher computational complexity both in the training and inference phases. Hence, the GNN is still not efficient.



\vspace{-4mm}
\subsection{Examples of the GNN Design}
\vspace{-0.1mm}

We first take the precoding optimization in a RIS system as an example to demonstrate how to design a GNN under the framework. Consider a multi-cell wireless system with $M$ BSs and one RIS with $N_{\sf RE}$ reflective elements (REs). Each BS is with $N_{\sf RF}$ RF chains and $N_{\sf B}$ antennas, and serves $K$ UEs each with a single antenna.

The baseband and analog precoding at BSs and the passive precoding at the RIS can be optimized, for example, to maximize the SE of the system under the maximal power constraint and modular constraints of the analog and passive precoding matrices (called problem \textsf{P-R}).

The precoding policy is the mapping from the environmental parameters to the optimized variables, where the environmental parameters are the channel matrices from the BSs to the UEs, from the BSs to the RIS, and from the RIS to the UEs.
The optimized variables are the baseband, analog and passive precoding matrices. Similar to problem \textsf{P-S} and \textsf{P-M}, the interference among users is not reflected in the environmental parameters but is reflected in the inner products of channel vectors and precoding vectors.

Using the method proposed in \cite{LSJ_MultiDim_GNN_2022}, we can identify four sets in the problem: UEs, RF chains (RFs), AN$^{\tt BS}$s and REs.
\begin{itemize}
	\item \emph{Type of sets}: The RE set is a normal set. The other three sets are nested sets, i.e., \\
	\{\{UEs in Cell$_1$\}, ..., \{UEs in Cell$_M$\}\}, \\
	\{\{RFs in Cell$_1$\}, ..., \{RFs in Cell$_M$\}\}, \\
	\{\{AN$^{\tt BS}$s in Cell$_1$\}, ..., \{AN$^{\tt BS}$s in Cell$_M$\}\}.
	\item \emph{Relation among sets}: The subsets in the UE set, AN$^{\tt BS}$ set and RF set need to be jointly permuted.
\end{itemize}



We next design the GNN.
In the $\ell$-th layer, the updated representation is an order-four tensor $\vec{\mathbf D}^{(\ell)}$. The update equation in the $\ell$-th layer is with four recursions, which are respectively used to satisfy the permutability to UEs, RFs, AN$^{\tt BS}$s and REs. The permutability to UEs is satisfied in the first recursion, because there exists interference among UEs.
\begin{enumerate}
	\item \emph{First recursion}: The I-O relation of the update equation should be a NPE function-II. This is because i) the UEs constitute a nested set, and ii) the ICI and MUI are not reflected in the environmental parameters of the problem.
	\item \emph{Second recursion}: The combiner and processor in the previous recursion should be NPE functions-I. This is because the RFs constitute a nested set,  and there is no interference among RFs.
        \item \emph{Third recursion}: The combiners and processors in the previous recursion should be NPE functions-I. This is because the AN$^{\tt BS}$s constitute a nested set,  and there is no interference among AN$^{\tt BS}$s.
        \item \emph{Fourth recursion}: The combiners and processors in the previous recursion should be APE functions-I. This is because the REs constitute a normal set, and there is no interference among REs. FNNs are designed as the processors and combiners of the APE functions.
\end{enumerate}

Due to the relation of the UE, AN$^{\tt BS}$and RF sets, only the elements in $\vec{\mathbf D}^{(\ell)}$ with the same cell index are updated.

In Table \ref{table:dnn-design}, we show how to design GNNs for precoding in different system settings.
 We only consider precoding policies, since the attention mechanism is needed according to our analyses. The GNN for learning other policies such as bandwidth and power allocation can be designed under the same framework.
\begin{table*}[!htb]
	\renewcommand{\arraystretch}{1.0}
	\setlength\tabcolsep{3pt}
	\footnotesize
	\caption{Efficient GNN Design for Several Wireless Problems}\label{table:dnn-design}
	\vspace{-2mm}
	\begin{tabular}{c|c|c|l|l|l|c|c}
		\hline\hline
		& \textbf{\begin{tabular}[c]{@{}c@{}}Num. of \\      recursions\end{tabular}} & \textbf{Sets}  & \multicolumn{1}{c|}{\textbf{Normal/Nested Set}}                                                                              & \multicolumn{1}{c|}{\textbf{\begin{tabular}[c]{@{}c@{}}One-set PE function  \\ in       each recursion\end{tabular}}} & \multicolumn{1}{c|}{\textbf{\begin{tabular}[c]{@{}c@{}}Independent/\\Joint\end{tabular}}}                                                                                                                                                                                    & \textbf{\begin{tabular}[c]{@{}c@{}}Whether an output\\function is cascaded\end{tabular}}                   & \textbf{Processor} \\ \hline
		\multirow{3}{*}{\textbf{\begin{tabular}[c]{@{}c@{}} Active and passive \\precoding in RIS-aided \\
					multi-cell narrowband\\ MU-MISO  system \end{tabular}}}   & \multirow{3}{*}{3}                                                          & UEs          & \begin{tabular}[c]{@{}l@{}}Nested:   cells and UEs\\       in each cell\end{tabular}                                      & \begin{tabular}[c]{@{}l@{}}NPE       function\end{tabular}                                           & \multirow{5}{*}{\begin{tabular}[c]{@{}l@{}}UE and AN$^{\tt BS}$ \\sets are joint, \\other sets are \\independent\end{tabular}}                                                                                                    & \multirow{3}{*}{\begin{tabular}[c]{@{}c@{}}Yes\end{tabular}}                                    & Attention          \\ \cline{3-5} \cline{8-8}
		&                                                                             & AN$^{\tt BS}$s    & \begin{tabular}[c]{@{}l@{}}Nested:   cells and AN$^{\tt BS}$s\\ in each cell\end{tabular}                                & \begin{tabular}[c]{@{}l@{}}NPE       function\end{tabular}                                           &                                                                                                                                                                                                                                        &                                                                                                                                     & Ordinary         \\ \cline{3-5} \cline{8-8}
		&                                                                             & REs   & Normal                                                                                                                   & APE   function                                                                                              &                                                                                                                                                                                                                                        &                                                                                                                                     & Ordinary         \\ \hline
		\multirow{4}{*}{\textbf{\begin{tabular}[c]{@{}c@{}} Active and passive \\ precoding in RIS-aided \\ single-cell narrowband \\      MU-MIMO system \end{tabular}}}  & \multirow{4}{*}{4}                                                          & DSs & \begin{tabular}[c]{@{}l@{}}Nested:   UEs and DSs\\ of each UE\end{tabular}                               & \begin{tabular}[c]{@{}l@{}}NPE       function\end{tabular}                                           & \multirow{4}{*}{\begin{tabular}[c]{@{}l@{}} AN$^{\tt UE}$ set\\ and DS set\\ are joint, other \\sets are \\independent\end{tabular}}                                                                                            & \multirow{4}{*}{\begin{tabular}[c]{@{}c@{}}Yes\end{tabular}}                                    & Attention          \\ \cline{3-5} \cline{8-8}
		&                                                                             & AN$^{\tt UE}$s  & \begin{tabular}[c]{@{}l@{}}Nested:   UEs and\\       AN$^{\tt UE}$s of each UE\end{tabular}                                   & \begin{tabular}[c]{@{}l@{}}NPE      function\end{tabular}                                           &                                                                                                                                                                                                                                        &                                                                                                                                     & Ordinary         \\ \cline{3-5} \cline{8-8}
		&                                                                             & AN$^{\tt BS}$s    & Normal                                                                                                                   & APE   function                                                                                              &                                                                                                                                                                                                                                        &                                                                                                                                     & Ordinary         \\ \cline{3-5} \cline{8-8}
		&                                                                             & REs   & Normal                                                                                                                   & APE   function                                                                                              &                                                                                                                                                                                                                                        &                                                                                                                                     & Ordinary         \\ \hline
		\multirow{4}{*}{\textbf{\begin{tabular}[c]{@{}c@{}} Baseband and analog \\precoding in single-cell \\   wideband    MU-MISO \\ system\end{tabular}}} & \multirow{4}{*}{4}                                                          & UEs          & Normal                                                                                                                  & APE   function                                                                                              & \multirow{4}{*}{Independent}                                                                                                                                                                                                           & \multirow{4}{*}{No}                                                                                                   & Attention          \\ \cline{3-5} \cline{8-8}
		&                                                                             & AN$^{\tt BS}$s    & Normal                                                                                                                  & APE   function                                                                                              &                                                                                                                                                                                                                                        &                                                                                                                                     & Ordinary         \\ \cline{3-5} \cline{8-8}
		&                                                                             & RFs      & Normal                                                                                                                   & APE   function                                                                                              &                                                                                                                                                                                                                                        &                                                                                                                                     & Ordinary         \\ \cline{3-5} \cline{8-8}
		&                                                                             & Sub-carriers   & Normal                                                                                                                  & APE   function                                                                                              &                                                                                                                                                                                                                                        &                                                                                                                                     & Ordinary         \\ \hline
		\multirow{9}{*}{\textbf{\begin{tabular}[c]{@{}c@{}} Baseband and analog\\ precoding in multi-cell\\wideband  MU-MIMO\\ system\end{tabular}}}          & \multirow{9}{*}{5}                                                          & DSs & \begin{tabular}[c]{@{}l@{}}Three-tier nested:   cells,\\ UEs in       each cell, and\\ DSs  of each UE\end{tabular}  & \begin{tabular}[c]{@{}l@{}}Three-tier NPE\\      function\end{tabular}                                           & \multirow{5}{*}{\begin{tabular}[c]{@{}l@{}}DS set,\\      AN$^{\tt UE}$ set, \\RF set, \\and AN$^{\tt UE}$ set\\ are joint, other \\sets are \\independent\end{tabular}} & \multirow{5}{*}{\begin{tabular}[c]{@{}c@{}}Yes\end{tabular}} & Attention          \\ \cline{3-5} \cline{8-8}
		&                                                                             & AN$^{\tt UE}$s  & \begin{tabular}[c]{@{}l@{}}Three-tier nested:   cells,\\ UEs in       each cell, and \\AN$^{\tt UE}$s of each UE\end{tabular} & \begin{tabular}[c]{@{}l@{}}Three-tier NPE\\      function\end{tabular}                                           &                                                                                                                                                                                                                                        &                                                                                                                                     & Ordinary         \\ \cline{3-5} \cline{8-8}
		&                                                                             & RFs      & \begin{tabular}[c]{@{}l@{}}Nested:   cells and RF\\       chains in each cell\end{tabular}                                  & \begin{tabular}[c]{@{}l@{}}NPE      function\end{tabular}                                           &                                                                                                                                                                                                                                        &                                                                                                                                     & Ordinary         \\ \cline{3-5} \cline{8-8}
		&                                                                             & AN$^{\tt BS}$s    & \begin{tabular}[c]{@{}l@{}}Nested:   cells and AN$^{\tt BS}$s\\  in each cell\end{tabular}                                & \begin{tabular}[c]{@{}l@{}}NPE      function\end{tabular}                                           &                                                                                                                                                                                                                                        &                                                                                                                                     & Ordinary        \\ \cline{3-5} \cline{8-8}
		&                                                                             & Sub-carriers   & Normal                                                                                                                   & APE   function                                                                                              &                                                                                                                                                                                                                                        &                                                                                                                                     & Ordinary         \\ \hline\hline
	\end{tabular}
	\vspace{-5mm}
\end{table*}

\vspace{-2.5mm}
\subsection{Complexity Analysis}\label{remark:complexity}
\vspace{-0.5mm}
Consider a problem with $S$ sets. There is interference in the dimension corresponding to one of the sets (denoted as the $S_{\sf interf}$-th set).
Using the same way of analyzing the computational complexity in the inference phase in \cite{GJ-TMLCN}, it is not hard to derive that the proposed GNN is with the number of floating-point operations (FLOPs) of $\mathcal{O}((N_{S_{\sf interf}})^2\prod_{s=1,s\neq S_{\sf interf}}^S N_s)$, where $N_s$ is the number of elements in the $s$-th set.

If we simply design the processors in all the one-set PE functions of a GNN as attention processors as in \cite{GJ-TMLCN}, the number of FLOPs of such a GNN is $\mathcal{O}(\prod_{s=1}^S (N_s)^2)$, which is $\prod_{s=1,s\neq S_{\sf interf}}^S N_s$ times higher than the GNN with judiciously designed processors in each recursion of the update equation.
For example, for learning precoding from problem \textsf{P-R} with $M=6$ cells, $N_{\sf B}=128$ antennas, $N_{\sf RF}=32$ RFs and $N_{\sf RE}=100$ REs, the number of FLOPs of a GNN with all attention processors is $MN_{\sf RF} \times MN_{\sf B} \times N_{\sf RE} \approx 1.5\times 10^7$ times higher than a GNN with the judiciously designed processors!

\vspace{-2mm}
\section{Simulation Results}\label{sec:simulation-results} \vspace{-1mm}
In this section, we validate the analyses by evaluating the performance of learning the hybrid precoding policy in RIS-aided MU-MISO system from problem \textsf{P-R}  with the proposed GNN.
We consider a single-cell downlink system, such that we can find an existing numerical algorithm as the performance baseline \cite{RIS_BCD}. The algorithm is used to solve the RIS-aided baseband precoding problem, which provides a performance upper bound of baseband and analog hybrid precoding.

We consider Rician channels  with the factor of 10, which consist of line-of-sight (LoS) and Non-LoS (NLoS) components. The path loss model is $32.6 + 36.7\log_{10}(d)$, where $d$ is the distance in meters. $P_{\max}=30$ dBm, $\sigma_0^2=-80$ dBm \cite{zbc_ris}.

The GNN is trained with unsupervised learning, where the negative SE over all the training samples is the loss function. Hence, each training sample includes only the inputs, which are the channel matrices generated from above-mentioned channel model. The learning
performance is measured by the SE ratio, i.e., the ratio of the SE achieved by the learned policy to the SE achieved by the block gradient descent (BCD) algorithm proposed in \cite{RIS_BCD}.

\begin{figure}[!htb]
\centering
\includegraphics[width=0.65\linewidth]{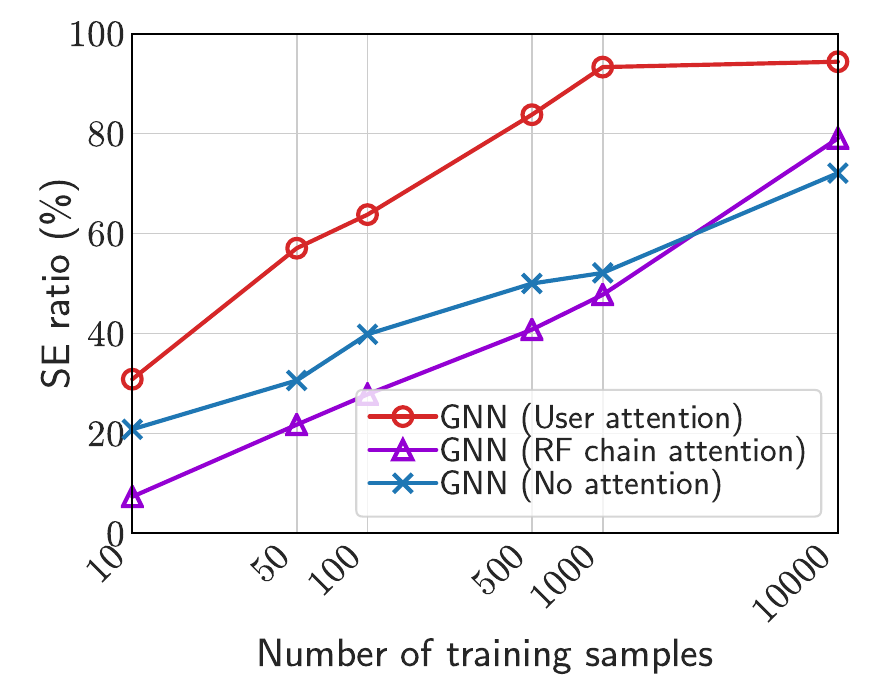}
\vspace{-1mm}
\caption{Learning performance, $N_{\sf B}=N_{\sf RE}=8$, $N_{\sf RF}=6$, $K=3$. The SE ratio of the BCD algorithm is 100\%.}
\label{fig:perf-ntr}
\vspace{-0.1mm}
\end{figure}

In Fig. \ref{fig:perf-ntr}, we show the impact of the attention processor on the learning performance with given samples, by comparing the SE ratios achieved by the following GNNs with recursive update equations. The problem scales (i.e., $K, N_{\sf RE}$ and $N_{\sf RF}$) are identical in training and test samples.

\begin{itemize}
	\item {\bf GNN (User attention)}, which is \emph{the proposed GNN} with attention processor on the user-dimension.
	\item \emph{GNN (RF attention)}, which is the GNN with attention processor only on the RF-dimension.
    \item \emph{GNN (No attention)}, which is the GNN with the ordinary processor on every dimension.

\end{itemize}

We have also evaluated the GNN using the attention processor only on the AN$^{\tt UE}$- or RE-dimension, whose performance is similar to \emph{GNN (RF-dimension)} and hence is not provided. All the GNNs are fine-tuned to achieve the best performance.
We consider a setting with a relatively small number of REs, because the proposed GNN can be generalized to a large value of $N_{\sf RE}$ (to be validated soon).

It can be seen  that \emph{\bf GNN (User attention)} achieves  a much higher SE ratio than the other two GNNs, meanwhile the sample complexity (i.e., the number of samples required to achieve an expected performance)  is much lower than the other two GNNs. This validates that an attention processor should be used in the dimension with interference.
The \emph{GNN (RF-dimension)} performs worse than \emph{GNN (No attention)} with a small number of training samples, because  more parameters need to be trained in the attention processor of \emph{GNN (RF-dimension)} than in the ordinary processor of \emph{GNN (No attention)}.

\begin{figure}[!htb]
\centering
\begin{minipage}[t]{0.48\linewidth}	
	\subfigure[Generalizability to $K$, $N_{\sf B}=8, N_{\sf RF}=6, N_{\sf RIS}=8$]{
		\includegraphics[width=\textwidth]{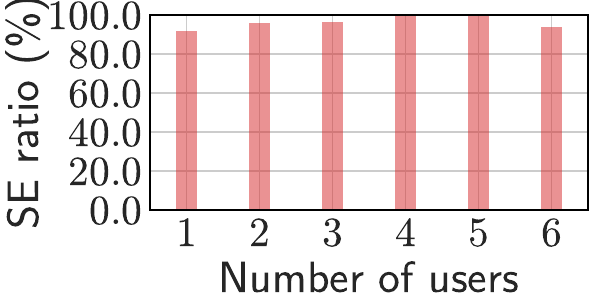}}
\end{minipage} \hspace{1mm}
\begin{minipage}[t]{0.48\linewidth}	
	\subfigure[Generalizability to $N_{\sf RF}$, $N_{\sf B}=16, N_{\sf RE}=40, K=3$]{
		\includegraphics[width=\textwidth]{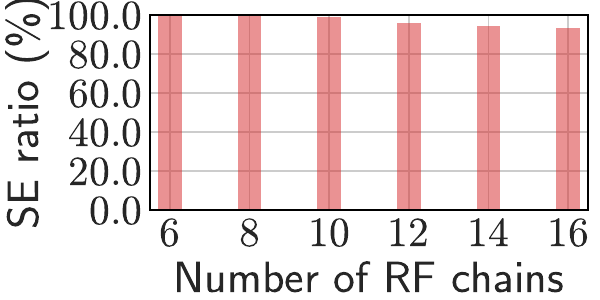}}
\end{minipage}

\begin{minipage}[t]{0.48\linewidth}	
	\subfigure[Generalizability to $N_{\sf B}$, $N_{\sf RF}=6,N_{\sf RE}=8,  K=3$]{
		\includegraphics[width=\textwidth]{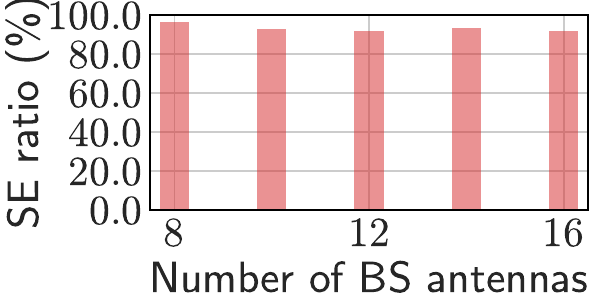}}
\end{minipage} \hspace{1mm}
\begin{minipage}[t]{0.48\linewidth}	
	\subfigure[Generalizability to $N_{\sf RE}$, $N_{\sf B}=8, N_{\sf RF}=6, K=3$]{
		\includegraphics[width=\textwidth]{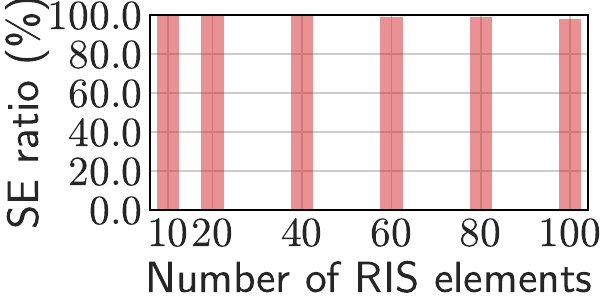}}
\end{minipage}
\vspace{-2mm}
\caption{Size generalizability of GNN (User attention).}\label{fig:perf}\vspace{-1mm}
\end{figure}

In Fig. \ref{fig:perf}, we show the impact of the attention processor on the size generalizability, by testing the \emph{\bf GNN (User attention)} in the scenarios with larger values of $K, N_{\sf RF}, N_{\sf RE}$ and $N_{\sf B}$ than the training samples without re-training.
The performances of the \emph{GNN (No attention)} and \emph{GNN (RF attention)} are not provided, because they cannot achieve satisfactory performance even when the problem scales are identical in training and test samples.
In each sub-figure, we only evaluate the generalizability to one of the three numbers while keeping the other numbers fixed.
When evaluating the generalization to $K$, the GNN is trained with 10,000 samples, which are generated in the scenarios where $K$ follows two-parameter exponential distribution with mean value of $2$ and standard deviation of $1$. In this case, $90$\% of samples are generated for the scenarios with $K\leq 3$. The test samples are generated in the scenarios with $K\thicksim{\mathbb U}(1,6)$, where ${\mathbb U}(\cdot,\cdot)$ stands for uniform distribution. When evaluating the generalization to $N_{\sf RF}$, $N_{\sf B}$, and $N_{\sf RE}$, the GNN is trained with 10,000 samples generated respectively in the scenarios with $N_{\sf RF}=6$, $N_{\sf B}=8$, and $N_{\sf RE}=10$. The test samples are generated in the scenarios with $N_{\sf RF}\thicksim{\mathbb U}(6,16)$, $N_{\sf B}\thicksim{\mathbb U}(8,16)$, and $N_{\sf RE}\thicksim{\mathbb U}(10,100)$, respectively. In sub-figure (b) for testing the generalizability to large numbers of RFs, the number of AN$^{\tt BS}$s and REs are set as larger values than the settings in the other three sub-figures, because $N_{\sf B} > N_{\sf RF}$, and $N_{\sf RE}$ is usually larger than $N_{\sf B}$. The value of $N_{\sf RE}$ is not set too large (say 100) such that the GNN can be trained without incurring an out-of-memory issue.
It can be seen that the SE ratio of the proposed GNN degrades slowly with the problem sizes, which indicates its good generalizability.

\vspace{-1mm}
\section{Conclusions}\label{sec:conclusions}
In this paper, we investigated how to effectively learn resource allocation policies by leveraging attention mechanism and permutation properties. We first identified the structures of the functions with high-dimensional permutation properties, by revealing the structures of several one-set PE functions. We then analyzed the structures of policies from several representative problems in interference-free and interference scenarios, from which we found that attention is only required in the dimensions with interference meanwhile the interference is not reflected in environmental parameters. Guided by these insights, we proposed a framework of designing efficient GNNs for resource allocation from the corresponding problems, where attention is only introduced in the necessary dimension. Simulation results demonstrated the good size generalizability and low training complexity of the designed GNN for learning the RIS-aided hybrid precoding policy.


\begin{appendices} \numberwithin{equation}{section}
\renewcommand{\thesectiondis}[2]{\Alph{section}}
\vspace{-2mm}
\section{~~~Proof of Proposition \ref{prop:APE}}\label{Appendix:re-express} \vspace{-1mm}
The function in  \eqref{eq:1d-pe-II} can be expressed as,
\begin{align}
	y_k \!&= g(x_k, [x_1,\cdots,x_{k-1},x_{k+1},\cdots,x_K])\notag\\
	\!&\overset{(a)}{=}\!g\Big(x_k, r\big(\textstyle\sum_{j=1,j\neq k}^K q_{\rm I}(x_j)\big)\Big) \!\overset{(b)}{=}\!f_{\rm I}(x_k, \textstyle\sum_{j=1,j\neq k}^K q_{\rm I}(x_j)), \notag
\end{align}
where $g(\cdot)$ is the composite function of $f_{\rm II}(\cdot)$ and $\sum_{j=1,j\neq k}^K$ $ q_{\rm II}(\cdot)$. In $(a)$, $[x_1,\!\cdots\!,x_{k-1},x_{k+1},\!\cdots\!,x_K]$ is expressed as a composite function $r\big(\textstyle\sum_{j=1,j\neq k}^K q_{\rm I}(x_j)\big)$. This can be expressed because the output of $q_{\rm I}(\cdot)$ is with dimension of $K$, such that a function $r(\cdot)$ can recover the vector $[x_1,\!\cdots\!,x_{k-1},x_{k+1},\!\cdots\!,x_K]$ from $\sum_{j=1,j\neq k}^K \!q_{\rm I}(x_j)$ \cite{zaheer2017deep}. In $(b)$, the composite of $g(\cdot)$ and $r(\cdot)$ is expressed as a function $f_{\rm I}(\cdot)$.

\vspace{-2mm}
\section{~~~Proof of Proposition \ref{prop:npe}}\label{appendix:proof-npe}
To prove this, we resort to the following lemmas that have been respectively proved in \cite{zaheer2017deep} and \cite{PE_decomposition}.

\begin{lemma}[\!\!\cite{zaheer2017deep}]\label{Cor:PI}
	Any function $y=\phi(\mathbf{x})$ that satisfies arbitrary permutation invariance property (API) $y=\phi(\mathbf{\Pi}^T\mathbf{x})$ can be expressed as
	$y = f(\sum_{j=1}^K q(x_j))$,
	where $\mathbf{x}\triangleq [x_1,\cdots,x_K]^T$, $f(\cdot):{\mathbb R}^{K+1}\mapsto{\mathbb R}$, and $q(\cdot):{\mathbb R}\mapsto{\mathbb R}^{K+1}$.
\end{lemma}

\begin{lemma}[\!\!\cite{PE_decomposition}]\label{Cor:PE}
	Any function $\mathbf{y}=\phi(\mathbf{x})$ that satisfies the APE property $\mathbf{\Pi}^T\mathbf{y}=\phi(\mathbf{\Pi}^T\mathbf{x})$ can be expressed as
	$y_k = f(x_k, \sum_{j=1, j\neq k}^K q(x_j))$,
	where $\mathbf{y}\triangleq [y_1,\cdots,y_K]^T, \mathbf{x}\triangleq [x_1,\cdots,x_K]^T$, $f(\cdot):{\mathbb R}^{K+1}\mapsto{\mathbb R}$, and $q(\cdot):{\mathbb R}\mapsto{\mathbb R}^K$.
\end{lemma}

The NPE property in Definition \ref{def:npe} is induced by both permutations of the sub-vectors and the elements in every sub-vector as follows.
\begin{enumerate}
	\item By regarding the sub-vectors in $\mathbf{x}$ and $\mathbf{y}$ as a whole, $\mathbf{y}=\phi(\mathbf{x})$ satisfies the APE property $\mathbf{\Pi}_0'\mathbf{y}=\phi(\mathbf{\Pi}_0'\mathbf{x})$, where $\mathbf{\Pi}_0'=\mathbf{\Pi}_0\otimes\mathbf{I}_K$.
	\item Denote the mapping from $\mathbf{x}$ to $\mathbf{y}_m$ as $\mathbf{y}_m=\phi_m(\mathbf{x})=\phi_m(\mathbf{x}_1,\cdots,\mathbf{x}_M)$. It satisfies $\mathbf{\Pi}_m^T\mathbf{y}_m=\phi_m(\mathbf{\Pi}_1^T\mathbf{x}_1,\cdots,\mathbf{\Pi}_M^T\mathbf{x}_M)$, i.e., it is APE to the elements in $\mathbf{x}_m$ and API to the elements in $\mathbf{x}_j, j=1,\cdots,M, j\neq m$.
\end{enumerate}

Considering the first PE property, according to Lemma \ref{Cor:PE}, $\mathbf{y}=\phi(\mathbf{x})$ can be expressed as,
\begin{equation}\label{eq:PE}
	\mathbf{y}_m = \textstyle f(\mathbf{x}_m, \sum_{j=1,j\neq m}^M q(\mathbf{x}_j)),
\end{equation}
where $f(\cdot): {\mathbb R}^{MK+K}\mapsto {\mathbb R}^K$, and $q(\cdot):{\mathbb R}^{K}\mapsto {\mathbb R}^{MK}$. The input and output dimensions of $f(\cdot)$, and the output dimension of $q(\cdot)$ are different from those in Lemma \ref{Cor:PE}, because $\mathbf{x}_m$ and $\mathbf{y}_m$ are vectors instead of scalars.

Next, we further consider the second permutation property.

Since $\mathbf{y}_k$ is APE to elements in $\mathbf{x}_k$, \!$y_{k_m}$ \!can \!be \!expressed \!as,
\begin{equation}\label{eq:PE2}
	\hspace{-0.1mm}y_{k_m}\!=\!f_{\rm I}\Big(x_{k_m}, \!\textstyle\sum_{j=1,j\neq k}^K q_{\rm I,1}(x_{j_m}), \!\sum_{j=1,j\neq m}^M q(\mathbf{x}_j)\Big),
\end{equation}
where $q_{\rm I,1}(\cdot):{\mathbb R}\mapsto{\mathbb R}^K$, $f_{\rm I}(\cdot): {\mathbb R}^{1+K+MK}\mapsto{\mathbb R}$.

Since $\mathbf{y}_k$ is API to elements in $\mathbf{x}_j, j=1,\cdots,K, j\neq k$, according to Lemma \ref{Cor:PI}, $q(\mathbf{x}_j)$ in \eqref{eq:PE2} can be expressed as $q(\mathbf{x}_j)=q_{\rm I,2}\big(\sum_{i=1}^K q_{\rm I,3}(x_{i_j})\big)$, where $q_{\rm I, 2}(\cdot):\! {\mathbb R}^{K+1}\!\mapsto\! {\mathbb R}^{MK}$, $q_{\rm I, 3}(\cdot):{\mathbb R}\mapsto {\mathbb R}^{K+1}$. The output of $q_{\rm I, 2}(\cdot)$ is a $MK$-dimensional vector instead of a vector as in Lemma \ref{Cor:PI}, because the output $q(\cdot)$ is a $MK$-dimensional vector.
By substituting $q(\mathbf{x}_j)=q_{\rm I,2}\big(\sum_{i=1}^K q_{\rm I,3}(x_{i_j})\big)$ into \eqref{eq:PE2}, we can obtain \eqref{eq:nest-pe-I}.

\vspace{-2mm}
\section{~~~Proof of Proposition \ref{prop:2d-pe}} \label{proof:2d-pe}
When the columns of $\mathbf{X}$ are permuted to $\mathbf{X}\mathbf{\Pi}_2=[\mathbf{x}_{\pi_2(1)},\cdots,\mathbf{x}_{\pi_2(K)}]$, where $[\pi_2(1),\cdots,\pi_2(K)]=[1,\cdots,K]\mathbf{\Pi}_2$, $\mathbf{y}_k$ becomes,
\begin{align}
	\mathbf{y}_k'&=f\big(\mathbf{x}_{\pi_2(k)}, \textstyle\sum_{j=1,j\neq k}^K q(\mathbf{x}_{\pi_2(j)})\big)\notag\\
	&=f\big(\mathbf{x}_{\pi_2(k)}, \textstyle\sum_{j=1}^K q(\mathbf{x}_{\pi_2(j)})-q(\mathbf{x}_{\pi_2(k)})\big)\notag\\
	&\overset{(a)}{=}f\big(\mathbf{x}_{\pi_2(k)}, \textstyle\sum_{j=1}^K q(\mathbf{x}_j)-q(\mathbf{x}_{\pi_2(k)})\big) \notag\\
	&=f\big(\mathbf{x}_{\pi_2(k)}, \textstyle\sum_{j=1,j\neq \pi_2(k)}^K q(\mathbf{x}_j)\big) \overset{(b)}{=} \mathbf{y}_{\pi_2(k)},
\end{align}
where $(a)$ holds because $\sum_{j=1}^K q(\cdot)$ is invariant to the permutations of the elements in $\mathbf{x}_1,\cdots,\mathbf{x}_K$, and $(b)$ comes from substituting $k=\pi_2(k)$ into \eqref{eq:2d-pe-r1}.

Hence, the columns of $\mathbf{Y}$ are permuted to $\mathbf{Y}\mathbf{\Pi}_2=[\mathbf{y}_{\pi_2(1)},\cdots,\mathbf{y}_{\pi_2(K)}]$. Then, $\mathbf{Y}\mathbf{\Pi}_2=\phi(\mathbf{X}\mathbf{\Pi}_2)$ holds. Similarly, we can prove from \eqref{eq:2d-pe-r2-1} and \eqref{eq:2d-pe-r2} that  $\mathbf{\Pi}_1^T\mathbf{Y}=\phi(\mathbf{\Pi}_1^T\mathbf{X})$ holds. Therefore, $\bm\Pi_1^T\mathbf{Y}\bm\Pi_2=\phi(\bm\Pi_1^T\mathbf{X}\bm\Pi_2)$ holds.

\vspace{-2mm}
\section{~~~Proof of Proposition \ref{prop:joint-pe}} \label{proof:joint-pe}
Since $\bm\Pi_1^T\phi(\mathbf{X})\bm\Pi_2=\phi(\bm\Pi_1^T\mathbf{X}\bm\Pi_2)$ holds for $\forall \bm\Pi_1, \bm\Pi_2$, we have $\bm\Pi^T\phi(\mathbf{X})\bm\Pi=\phi(\bm\Pi^T\mathbf{X}\bm\Pi)$. Moreover, it is not hard to prove that $\mathbf{I}_K=\bm\Pi^T\mathbf{I}_K\bm\Pi$. Hence, $\bm\Pi^T\mathbf{Y}\bm\Pi=\bm\Pi^T(\phi(\mathbf{X})\odot\mathbf{I}_K + \mathbf{X}\odot (1-\mathbf{I}_K))\bm\Pi=(\bm\Pi^T\phi(\mathbf{X})\bm\Pi)\odot(\bm\Pi^T\mathbf{I}_K\bm\Pi) + (\bm\Pi^T\mathbf{X}\bm\Pi)\odot(1-\bm\Pi^T\mathbf{I}_K\bm\Pi)=\phi(\bm\Pi^T\mathbf{X}\bm\Pi)\odot \mathbf{I}_K+(\bm\Pi^T\mathbf{X}\bm\Pi)\odot (1-\mathbf{I}_K)$.

\vspace{-2mm}
\section{~~~Proof of Proposition \ref{prop:reexp-power-bandw-allo}}\vspace{-1mm}
\label{proof:prop:reexp-power-bandw-allo}
To prove this proposition, we find the summations in each iteration that can be regarded as information aggregation with $\sum_{j=1,j\neq k}^K q_{\sf B}(\cdot)$. Then, we re-express other operations except the summations as information combination with $f_{\sf B}(\cdot)$.

To re-express the iterative equations of the algorithm as \eqref{eq:pb}, we add an auxiliary equation,
\begin{equation}\label{eq:aux}
	g_k = g_k,
\end{equation}
and define a vector $\mathbf{d}_k^{(\ell)} \triangleq [d_{k,1}^{(\ell)},d_{k,2}^{(\ell)},d_{k,3}^{(\ell)},d_{k,4}^{(\ell)},d_{k,5}^{(\ell)}]^T=[p_k^{(\ell)},B_k^{(\ell)},\mu^{(\ell)},\lambda^{(\ell)},g_k]^T$.

(i) Finding the aggregator: There is only one summation in the iterative equations, i.e., in \eqref{eq:iter-bp-4}. To show that it plays the role of information aggregation, we express each term (say the $j$-th term) in the summation as a function of $\mathbf{d}_j^{(\ell)}$,
\begin{equation}\label{eq:qb-expression}
	p_j^{(\ell)} = d_{j,1}^{(\ell)} = [1,0,0,0,0]\cdot\mathbf{d}_j^{(\ell)}\triangleq q_{\sf B}(\mathbf{d}_j^{(\ell)}).
\end{equation}
Then, the summation in \eqref{eq:iter-bp-4} can be written as
$e_k^{(\ell)} = \textstyle\sum_{j=1}^K q_{\sf B}(\mathbf{d}_j^{(\ell)})$.

(ii) Finding the combiner: With the expressions of $\mathbf{d}_k^{(\ell)}$ and $e_k^{(\ell)}$, we can re-write \eqref{eq:iter-bp-1}$\sim$\eqref{eq:iter-bp-4} and \eqref{eq:aux} in compact form as follows,
\begin{align}\label{eq:fb-expression}
	&\mathbf{d}_k^{(\ell+1)}=[p_k^{(\ell+1)},B_k^{(\ell+1)},\mu_k^{(\ell+1)},\lambda^{(\ell+1)},g_k]^T \notag\\
	&=\textstyle\Big[\Big[p_k^{(\ell)}+\frac{\mu_k^{(\ell)}g_k}{\ln 2}\frac{1}{N_0B_k+p_k^{(\ell)}g_k}-\lambda^{(\ell)}\Big]^+,\notag\\
	&\textstyle\Big[B_k^{(\ell)}+1+\mu_k^{(\ell)}\log_2\Big(1+\frac{p_k^{(\ell)}g_k}{N_0B_k^{(\ell)}}\Big)-\mu_k^{(\ell)}\frac{p_k^{(\ell)}g_k}{B_k^{(\ell)}N_0+p_k^{(\ell)}g_k}\Big]^+,\notag\\
	&\textstyle \Big[\mu_k^{(\ell)} + B_k^{(\ell)}\log_2(1+p_k^{(\ell)}g_k/(N_0B_k^{(\ell)}))-s_0\Big]^+,
	\notag\\
	&\textstyle \Big[\lambda^{(\ell)} + e_k^{(\ell)}-P_{\max}\Big]^+, g_k\Big]\notag\\
	&\triangleq f_{\mathsf{B}}'\big(\mathbf{d}_k^{(\ell)}, e_k^{(\ell)}\big)= f_{\mathsf{B}}'\big(\mathbf{d}_k^{(\ell)}, \textstyle\sum_{j=1}^K q_{\mathsf{B}}(\mathbf{d}_j^{(\ell)})\big) \notag\\
	& = f_{\mathsf{B}}\big(\mathbf{d}_k^{(\ell)}, \textstyle\sum_{j=1,j\neq k}^K q_{\mathsf{B}}(\mathbf{d}_j^{(\ell)})\big).
\end{align}
The I-O relation of the compact form can be expressed as a function $f_{\mathsf{B}}'(\mathbf{d}_k^{(\ell)}, e_k^{(\ell)})$ because all of the variables except for $e_k^{(\ell)}$, i.e., $p_k^{(\ell)},B_k^{(\ell)},\mu_k^{(\ell)},\lambda^{(\ell)},g_k$, are elements in $\mathbf{d}_k^{(\ell)}$. $P_{\max}, s_0$ and $N_0$ are regarded as constants.
\end{appendices}
\vspace{-3mm}
\begin{spacing}{1}
\bibliography{IEEEabrv,GJ}
\end{spacing}
\end{document}